\documentclass[12pt]{article}
\pdfoutput=1

\usepackage{tikz}

\usepackage{amsmath}

\usepackage[
      colorlinks=true,
      linkcolor=blue,
      urlcolor=blue,
      filecolor=black,
      citecolor=red,
      pdfstartview=FitV,
      pdftitle={},
        pdfauthor={Benjamin Withers},
        pdfsubject={},
        pdfkeywords={},
        pdfpagemode=None,
        bookmarksopen=true
      ]{hyperref}

\newcommand{\be}{\begin{equation}}
\newcommand{\ee}{\end{equation}}
\newcommand{\ba}{\begin{eqnarray}}
\newcommand{\ea}{\end{eqnarray}}

\begin{document}

\makeatletter
\renewcommand{\theequation}{\thesection.\arabic{equation}}
\@addtoreset{equation}{section}
\makeatother

\baselineskip 18pt

\begin{titlepage}

\vfill

\begin{flushright}
\end{flushright}

\vfill

\begin{center}
   \baselineskip=16pt
   {\Large\bf Holographic Checkerboards}
  \vskip 1.5cm
      Benjamin Withers\\
   \vskip .6cm
   b.s.withers@soton.ac.uk\\
     \vskip .6cm
     \textit{Mathematical Sciences and STAG Research Centre,\\
University of Southampton,
Highfield, Southampton SO17 1BJ, UK}


\end{center}

\vfill

\begin{center}
\textbf{Abstract}
\end{center}

\begin{quote}
We construct cohomogeneity-three, finite temperature stationary black brane solutions dual to a field theory exhibiting checkerboard order. The checkerboards form a backreacted part of the bulk solution, and are obtained numerically from the coupled Einstein-Maxwell-scalar PDE system. They arise spontaneously and without the inclusion of an explicit lattice. The phase exhibits both charge and global $U(1)$-current modulation, which are periodic in two spatial directions. The current circulates within each checkerboard plaquette. We explore the competition with striped phases, finding first-order  checkerboard to stripe phase transitions.\newline
We also detail spatially modulated instabilities of asymptotically AdS black brane backgrounds with neutral scalar profiles, including those with an hyperscaling violating IR geometry at zero temperature.
\end{quote}

\vfill

\end{titlepage}

\tableofcontents

\section{Introduction}
The use of holography to describe strongly coupled gauge theories, including models for metallic, insulating and superconducting phases has received a great deal of attention recently.  A variety of metals exhibit phases in which translational invariance or the underlying lattice symmetries are spontaneously broken, see \cite{2009AdPhy..58..699V} for a review. One possible outcome is striped order where translations are only broken in one direction resulting in a periodic configuration. Another possibility is checkerboard order, where no continuous translational symmetry remains and the phase becomes periodic in two spatial directions.

The spontaneous breaking of spatial symmetries can be modelled using holography, using a variety of holographic models \cite{Donos:2011bh, Nakamura:2009tf, Donos:2011pn, Donos:2011qt, Domokos:2007kt, Ooguri:2010xs, Bayona:2011ab, Bergman:2011rf}. Charge density may be introduced by turning on a chemical potential which sources a $U(1)$ gauge field in the bulk, and a much used gravitational model which provides these minimal ingredients is Einstein-Maxwell theory, admitting asymptotically AdS Reissner-Nordstrom black brane (RN) solutions describing the normal phase. In 4d the spatial symmetries of the $R^2$ brane directions of RN may be spontaneously broken. 

A symmetry breaking scenario of this type occurs in a Einstein-Maxwell-(pseudo)scalar model,
\be
S = \int d^4x \sqrt{-g}\left(R - \frac{1}{2}(\partial\phi)^2 - \frac{\tau(\phi)}{4}F^2 - V(\phi)\right) - \int \frac{\vartheta(\phi)}{2}F\wedge F \label{action}.
\ee
where we set $16\pi G =1$.
In \cite{Donos:2011bh} a perturbative analysis about the RN solution revealed marginally unstable modes breaking translational, as well as $P$ and $T$ symmetries.\footnote{Instabilities which do not break P and T can also exist within this general class of models, without the $\vartheta$ term. This has been shown by a near-horizon, AdS$_2\times R^2$ analysis at $T=0$ in  \cite{Donos:2013gda}.}
The perturbations contain charge density waves and, due to the parity violating term in \eqref{action}, also include a spatially modulated current. These marginal modes indicate the existence of a corresponding branch of black brane solutions which break translations. 

One possibility is that the branch is spatially modulated in only one direction, referred to as striped black branes, dual to the striped phases of the field theory. Solutions of this type have been constructed \cite{Donos:2013wia, Withers:2013loa, Rozali:2012es, Withers:2013kva,Rozali:2013ama}, and are obtained numerically, solving a set of nonlinear elliptic 2d PDEs resulting from the bulk Einstein-matter equations. In \cite{Withers:2013kva} the space of striped solutions was constructed with the dominant stripe's wavelength dependent on the temperature. It was also shown numerically that the line of dominant stripe solutions obeyed a particular averaged thermodynamic relation, which was subsequently derived as the result of a first-law relation involving the periodicity of the solution \cite{Donos:2013cka}. 

Another possibility is that there are black brane solutions which break translations in both boundary spatial directions. At least, at the linear level the marginal modes may simply be rotated and superposed. It is also possible that the resulting inhomogeneous black brane solutions are thermodynamically dominant to the striped solutions.  The purpose of this paper is to seek these cohomogeneity-three black brane solutions with no surviving continuous translational invariance, which we shall refer to as checkerboard phases. We explicitly construct backreacted checkerboard configurations and explore their competition with striped phases. We find normal-to-checkerboard and checkerboard-to-stripe first order phase transitions as the temperature is lowered.\footnote{There are further possibilities of course, such as a triangular lattice. We have not investigated such possibilities in this work, though they would be a interesting future direction.}

Finally, returning to the instability analysis \cite{Donos:2011bh}, the linear marginally unstable modes of the RN solution at wavenumber $k$ involve a spatially modulated current for the global $U(1)$, $\left<J_y\right> \propto \cos{k x}$. However, at linear order they do not involve a modulation of the charge density, $\left<J_t\right>$. A charge density wave does occur at higher order in perturbation theory and with half the period, i.e. $\left<J_t\right> - \bar{J}_t \propto \cos{2k x}$ where $\bar{J}_t$ denotes a homogeneous component. In this paper we extend these instabilities to include deformations by the operator dual to $\phi$. This is achieved by setting a constant source in the near-boundary expansion for $\phi$.  This changes the nature of the instability; whilst it still breaks $P$ and $T$, the charge density is modulated at leading order in a perturbative instability analysis and with wavenumber $k$.

The paper is arranged as follows. In section \ref{setup} we outline the specific model choice, discuss the numerical method, boundary conditions and associated renormalised one-point functions of the dual theory. In section \ref{prelim} we present a checkerboard solution for the model investigated in \cite{Withers:2013loa} and investigate the space of rectangular checkerboards at fixed temperature. In section \ref{p1def} we discuss spatially modulated instabilities with $O_\phi$ deformations. In section \ref{transition} we present our main results for the deformed model -- the existence of a dominant line of checkerboard solutions, reached by first order transition from the normal phase, and in some cases a first order transition to the striped phase at lower temperatures. We conclude in section \ref{comments}.  

\section{Setup \label{setup}}
We make the following choices for the model \eqref{action},
\be
\tau(\phi) = \text{sech}{\frac{\sqrt{n}}{2\sqrt{3}}\phi},\quad V(\phi) = -6 \cosh{\frac{\phi}{\sqrt{3}}},\quad \vartheta(\phi) = \frac{c_1}{6\sqrt{2}} \tanh{\sqrt{3}\phi} \label{taudef}
\ee
introducing the parameters $(n,c_1)$ corresponding to those used in the linear instability analysis of \cite{Donos:2011bh}. In all cases $\phi$ is dual to an operator of dimension $\Delta =  2$ with a $\Delta = 1$ source, $\phi^{(1)}$ which appears as the leading piece in a near boundary expansion:
\be
\phi(z,x^\mu) = \phi^{(1)} z + \phi^{(2)}(x^\mu)z^2+O(z)^{3}. \label{phibdy}
\ee
We shall consider checkerboards in the presence of \emph{homogeneous} deformations of this type and, as we shall see, it facilitates the competition of checkerboard phases with striped phases for the models considered.

The case $n=36,c_1 = 6\sqrt{2}$ arises in a consistent truncation from M-theory on an arbitrary SE$_7$ space \cite{Gauntlett:2009zw}, and motivates the functional form of \eqref{taudef}.  This particular choice has been well-studied, utilised in the study of holographic superconductors \cite{Gauntlett:2009bh} and striped instabilities \cite{Donos:2011bh}, including competition with superconductors under the influence of a magnetic field \cite{Donos:2012yu}. In \cite{Withers:2013loa} this model was used to construct striped black branes. For these parameters the critical temperature for the striped instability is particularly low, consequentially a small adjustment to the model was made, setting $n=36, c_1=9.9$  to improve the critical temperature. In section \ref{prelim} we construct checkerboard black branes using this model; we find that only striped phases dominate, though we have not performed an exhaustive analysis. In section \ref{transition} we study the case $n=0,c_1=9.9$, where we find that the checkerboard phases dominate with first order phase transitions once deformed by turning on $\phi_{(1)}$.

\subsection{Gauge fixing and numerical method\label{gaugefix}}

We will directly construct stationary solutions of the coupled Einstein-Maxwell-scalar equations. Without modification or gauge fixing this is not an elliptic pde system as we require for this boundary value problem. An elliptic system can be reached via a suitable modification of the equations using the DeTurck method \cite{Headrick:2009pv,Adam:2011dn,Wiseman:2011by}. For the Einstein equations the Ricci tensor is replaced by
\be
R_{MN} \to R^{H}_{MN} = R_{MN} - \nabla_{(M} \xi_{N)}
\ee
where we have introduced the DeTurck vector $ \xi^M = g^{NP}(\Gamma_{NP}^M - \tilde\Gamma_{NP}^{M})$
and where $\tilde{\Gamma}$ is the Christoffel symbol for a reference metric $\tilde{g}_{ab}$. We demand that $\xi$ vanish on a solution of the modified equations so that we also have a solution to the unmodified system. In addition we find it necessary to remove longitudinal gauge modes of the $U(1)$ gauge field $A$. A sufficient modification for this purpose is,
\be
\nabla_M F^{M}_{~N} \to \left(\nabla_M F^{M}_{~N} \right)^{H} = \nabla_a F^{M}_{~N} + \partial_N \psi
\ee
where we have defined the scalar $\psi = \tilde{g}^{MN} \tilde{\nabla}_M (A_N - \tilde{A}_N)$ with reference gauge field $\tilde{A}_M$. With this additional term the principle part of gauge field equation no longer vanishes for longitudinal/gauge modes.\footnote{We could also have constructed the scalar $\psi$ using the metric $g$ making the gauge field analysis cleaner, though note this introduces new second derivative terms for the metric. We find that the present definition of $\psi$ works well in practise.} In addition to $\xi^M$, we now require that the scalar $\psi$ vanishes on any solutions. 

The resulting set of  Einstein-Maxwell-scalar equations with the additional terms are then solved numerically using a Newton-Raphson method. We use a grid of size $N^3 = N_z\times N_x\times N_y$, represented spectrally with Fourier modes in $x,y$, the spatial boundary directions and Chebyschev polynomials in $z$, the radial holographic direction. Solutions are periodic in the $x$ and $y$ directions with wavenumbers $k_x$ and $k_y$ respectively. For lower grid sizes we find it convenient to take $N_z=2N_x=2N_y$,  primarily as it assists in the accurate extraction of the free energy from the numerical data. We will provide convergence checks of $\xi$ and $\psi$ and the free energy with $N$. 

Finally, some comments on implementation. 
With the inclusion of currents resulting from $P$ and $T$ breaking through the $\vartheta F\wedge F$-term, the system studied has 15 equations of motion (for 15 fields: 10 metric, 4 gauge field and 1 scalar), as described in the next section, \ref{ansatzsec}. We find it convenient to generate these only numerically, at runtime. 
Then, each step in the Newton-Raphson iterative method consists of two distinct computational stages: first constructing the Jacobian matrix of first derivatives of the equations (including boundary conditions) with respect to the fields. This may be done with a numerical finite differencing and may be efficiently parallelised. Because it is a 3d problem containing at most second derivatives, the resulting matrix is sparse despite being spectrally represented. The second Newton-Raphson stage is solving the resulting sparse linear system, for which we use a bi-conjugate gradient method. Good convergence occurs within only a handful of Newton-Raphson steps.

\subsection{Ansatz and boundary conditions\label{ansatzsec}}
To start we can consider a metric ansatz for a coordinate system adapted to the timelike Killing vector field $T=\partial_t$, 
\be
ds^2 = g_{MN}dx^Mdx^N= g_{tt}(x^c) (dt + \mathcal{A}_a(x^c) dx^a)^2 + h_{ab}(x^c) dx^a dx^b
\ee
where $a=1,2,3$ with $x^a = (z,x,y)^a$. The horizon position dictated by $g_{tt}(x_h)=0$ where $T$ is null. We do not restrict the base manifold with metric $h$ to have any particular symmetries.  Similarly, $\mathcal{A}_a$, will be non-zero in general. In the absence of gauge fixing this ansatz is then just a way of packaging the full quota of 10 metric functions, each of which is $t$-independent but depend on the three remaining $x^a$ coordinates. In practise we repackage the above ansatz for convenience:
\begin{eqnarray}
g_{zz} = \frac{1}{z^2 f(z)} G_{zz},\qquad g_{tt} = -\frac{1}{z^2} f(z) G_{tt},\qquad g_{ij} = \frac{1}{z^2} G_{ij}\\
g_{ti} = \frac{f(z)}{z^2} G_{ti},\qquad g_{tz} = f(z) G_{tz},\qquad g_{zi} = G_{zi}
\end{eqnarray}
where $i=1,2$ labels boundary spatial directions, $x^i=(x,y)^i$. Each of $G_{MN}$ are functions of $(z,x,y)$ and $f(z) = \frac{1}{4}(4+ \mu^2 z^4 -z^3(4+\mu^2))$, with $f(z_h) = 0$ giving the horizon location, chosen to lie at $z = z_h=1$.

When $G_{zz} = G_{tt} = G_{xx} = G_{yy} = 1$, $G_{xy} = G_{ti} = G_{tz} = G_{zi} = 0$, $\phi = A_z = A_i = 0$ and $A_t = \mu (1-z)$ we recover the RN solution. The reference metric $\tilde g$ and reference gauge field $\tilde A$ are chosen to take on these values.

\subsubsection{UV}
In the UV near $z=0$ for solutions where $\xi = \psi = 0$ we have the following expansions for the matter fields
\ba
\phi(z,x^i) &=& \phi^{(1)} z + \phi^{(2)}(x^i)z^2+O(z)^3\\
A_t(z,x^i) &=& \mu + A_t^{(1)}(x^i)z + O(z)^2\\
A_z(z,x^i) &=& O(z)^2\\
A_{i}(z,x^i) &=& A^{(1)}_{i}(x^i) z+ O(z)^2 
\ea
and for the metric functions,
\ba
G_{zz}(z,x^i) &=& \frac{1}{3}\phi_{(1)}\phi_{(2)}(x^i)z^3+ O(z)^4 \\
G_{\mu\nu}(z,x^i) &=& \delta_{\mu\nu}-\frac{\phi_{(1)}^2}{8}\delta_{\mu\nu}\,z^2 + G_{\mu\nu}^{(3)}(x^i)z^3+ O(z)^4\\
G_{z\mu}(z,x^i) &=& O(z)^2 
\ea
with additional relations constraining subleading data such that they satisfy U(1)-current conservation, stress tensor conservation and a conformal Ward identity, derived in section \ref{1ptfcns}. From these expansions we can simply read off the Dirichlet boundary conditions required for the numerics. In the case of $\phi$ we work with $\phi_c(z,x^i) \equiv \phi(z,x^i)/z$ and set a (constant) Dirichlet condition for it in the UV.
\subsubsection{Horizon}
At $z=1$ we seek a regular, non-extremal horizon. This entails a near-$z=1$ expansion for the matter fields,
\ba
\phi(z,x^i) &=& \phi_{+}(x^i) + O(1-z)\\
A_t(z,x^i) &=& E_{+}(x^i) (1-z) + O(1-z)^2\\
A_z(z,x^i) &=&  \frac{4}{12-\mu^2}\partial_i A_{i}^+(x^i)+O(1-z)\label{AzHorizon}\\
A_{i}(z,x^i) &=& A_{i}^{+}(x^i) + O(1-z)
\ea
where the behaviour of $A_z$ is determined by the near horizon expansion of the $\psi =0$ condition. For the metric functions, 
\ba
G_{zz}(z,x^i) &=&  G_{zz}^+(x^i) + O(1-z)\\
G_{\mu\nu}(z,x^i) &=& G_{\mu\nu}^{+}(x^i) + O(1-z)\\
G_{z\mu}(z,x^i) &=& G_{z\mu}^{+}(x^i) + O(1-z)
\ea
with a Dirichlet boundary condition $G_{zz}^{+} = G_{tt}^{+}$. Subleading orders in this expansion are determined by these leading data.  We use a second Dirichlet condition for $A_t(z,x^i) = 0$.  For the remaining fields we impose the leading form of the expansion. In particular, defining the quantity $\Pi_\phi \equiv \partial_z ((1-z) \phi(z,x^i))$, the boundary condition used is $\Pi_\phi = -\phi$ at $z=1$. As a check of this boundary condition, of the near-horizon behaviour and as a more general check of the solutions, we confirm that the analytic relations between the coefficients of the $O(1-z)$ terms and leading order data hold numerically on the solutions constructed. Similarly the leading part of \eqref{AzHorizon} holds with the above boundary condition, or alternatively may be fixed using a Dirichlet condition giving equivalent results.

Finally, the temperature is given by
\be
T = \frac{-f'(z)}{4\pi}\bigg|_{z=1} = \frac{12-\mu^2}{16\pi}, \label{temperature}
\ee
and the entropy density by
\be
s(x^i) = \frac{a(x^i)}{4G} = 4\pi a(x^i) = 4\pi \sqrt{G_{xx}^{+}G_{yy}^{+} - (G_{xy}^{+})^2}.\label{localentropy}
\ee

\subsection{One-point functions\label{1ptfcns}}
To perform renormalisation of the action \eqref{action} and extract the one-point functions we first transform the metric to Fefferman-Graham form near the AdS boundary. Details are provided in appendix \ref{apprenorm}. Here we will be brief and just quote the key results. The renormalised action including the appropriate Gibbons-Hawking term is given by
\be
S_{ren} = S - 2\int d^3x \sqrt{-\gamma} K + 2 \int d^3x \sqrt{-\gamma}\left(-2 - \frac{1}{4}\phi^2\right).
\ee
In total the one point functions are summarised by the variation, 
\be
\delta S_{ren} = \int d^3x \sqrt{-\gamma_{(0)}}\left(\frac{1}{2}\left<T^{\mu\nu}\right>\delta \gamma^{(0)}_{\mu\nu} + \left<O_\phi\right>\delta\phi_{(1)}+ \left<J^\mu \right> \delta A^{(0)}_\mu\right).
\ee
where $x^\mu = (t,x,y)^\mu$,  $\gamma$ is the induced metric on the boundary, and sub/superscripts in brackets label coefficients in the small-$z$ near boundary expansion. For instance, $\phi_{(1)}$ is as defined in \eqref{phibdy}. The one point functions are given by,
\begin{eqnarray}
\left<T^{\mu\nu}\right>  &=&  3 \left(g_{(3)}^{\mu\nu} - g_{(0)}^{\mu\nu} \text{tr} g^{(3)} - \frac{2}{3} g_{(0)}^{\mu\nu} g^{(3)}_{zz}-\frac{1}{3} g_{(0)}^{\mu\nu}\phi_{(1)}\phi_{(2)}\right)\\
\left<O_\phi\right> &=&   \phi_{(2)}\\
\left<J^\mu\right> &=& A_{(1)}^\mu
\end{eqnarray}
In addition invariance under the Weyl transformations $\delta \gamma^{(0)}_{\mu\nu} = -2 \lambda \gamma^{(0)}_{\mu\nu}$, $\delta \phi_{(1)} = \lambda \phi_{(1)}$ gives the conformal Ward identity,
\be
\left<T^\mu_\mu\right> =  \phi_{(1)}\left<O_\phi\right>.
\ee
Inserting the expressions for the one-point functions above we verify this holds using the equations of motion. We can also use this identity as a check of the numerical extraction of the relevant stress tensor components.

An expression for the free energy density is given by,
\be
w(x^i) = \left<T^{tt}\right>(x^i) - T s(x^i) - \mu \left<J^t\right>(x^i).
\ee
where $T$ and $s(x^i)$ are defined in \eqref{temperature} and \eqref{localentropy}.
Throughout we will adopt thermodynamical quantities averaged over the spatial boundary directions of the computational domain, these will be denoted with a bar, for instance,
\be
\bar{\omega} = \frac{k_x}{2\pi} \frac{k_y}{2\pi} \int_{-\pi/k_x}^{\pi/k_x} dx \int_{-\pi/k_y}^{\pi/k_y}  dy\; \omega(x,y) . 
\ee
\newline\newline Finally note we work in the grand canonical ensemble and appropriately construct dimensionless quantities using powers of the chemical potential, $\mu$. This allows us to set $\mu =1$ where it streamlines presentation, specifically we shall do this where we present numerical results.

\section{Checkerboards\label{prelim}}
In this section we use the model \eqref{action}, \eqref{taudef} at $n=36,c_1 = 9.9$ and with no scalar source, $\phi^{(1)} = 0$. The dominant striped phases for this case were constructed in \cite{Withers:2013kva}. First consider the linear instability of RN, details of which can be found in \cite{Donos:2011bh,Withers:2013kva}. This is formed from the following set of consistent perturbations with momentum $k$,
\be
\delta\phi = \lambda(z) \cos{kx}, \qquad \delta A_y = a_{y} (z)\sin{kx},\qquad   \delta g_{ty} = h_{ty}(z) \sin{kx} \label{instabnodef}
\ee
subject to horizon regularity and normalisability at the boundary. This results in a `bell-curve' of k-dependent critical temperatures. At higher orders the charge density becomes modulated with period $2k$. In what follows we study the nonlinear branch of black brane solutions which emerges from a linear combination of two such modes, one modulated in the $x$-direction as above, and one modulated in the $y$-direction.

The solution is a checkerboard, periodic in $x$ and $y$ with momenta $k_x$ and $k_y$ respectively. The highest temperature at which the checkerboards connect with RN is the same as for the striped solutions and is given by $T=T^\ast = 0.0236$ for a square configuration $k_x=k_y=k^*= 0.783$. Fixing for now $k_x=k_y=k^\ast$ the checkerboard phase at $T=0.55 T^\ast$ is presented below.

\begin{figure}[h!]
\begin{center}
\includegraphics[width=0.35\columnwidth]{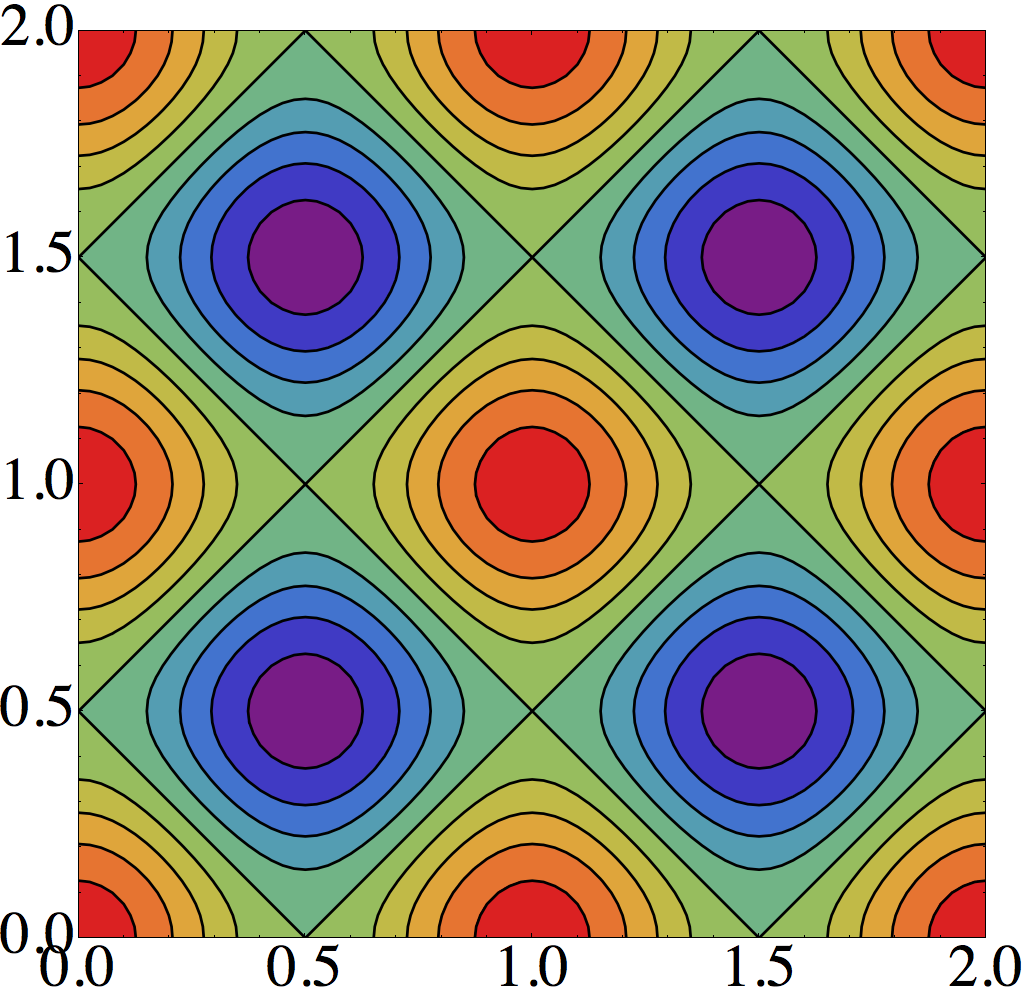}
\hspace{40pt}
\includegraphics[width=0.35\columnwidth]{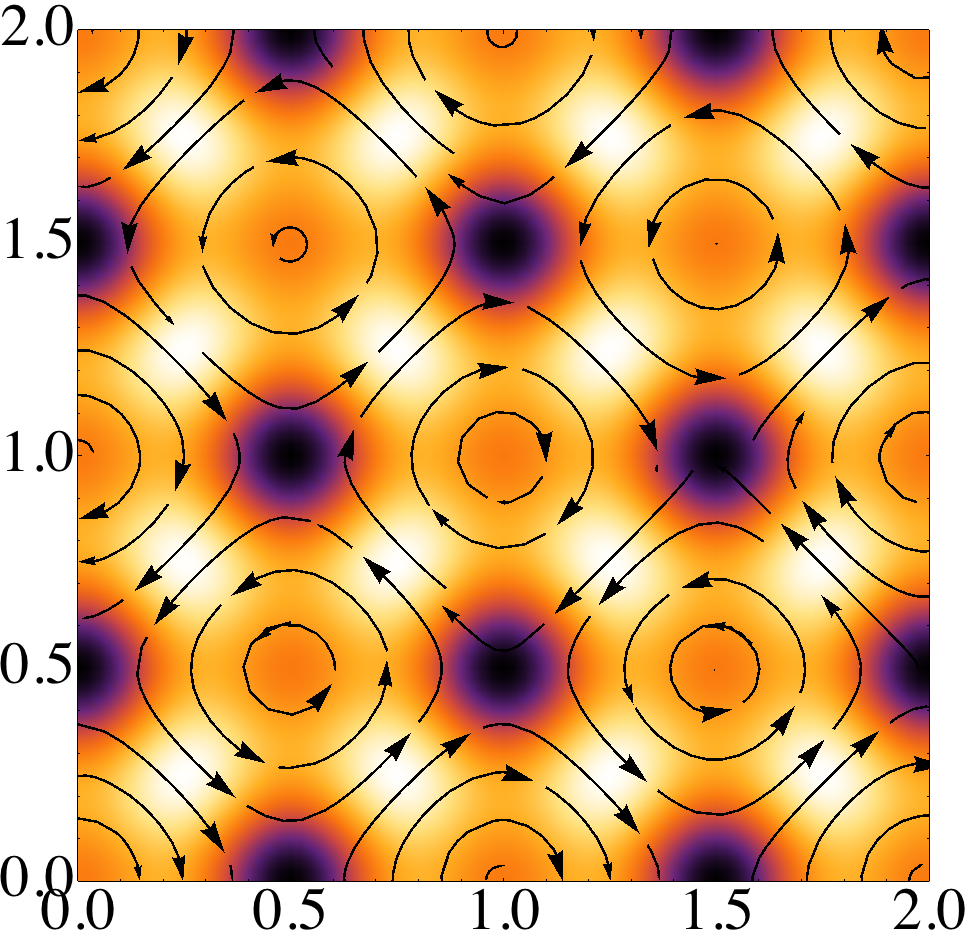}
\begin{picture}(0.1,0.1)(0,0)
\put(-380,80){\makebox(0,0){$\frac{k_y }{2\pi} y$}}
\put(-80,-10){\makebox(0,0){$k_x x / 2\pi$}}
\put(-175,80){\makebox(0,0){$\frac{k_y}{2\pi} y$}}
\put(-285,-10){\makebox(0,0){$k_x x / 2\pi$}}
\end{picture}
\vskip 1em
\caption{A holographic checkerboard in the $n=36,c_1 =9.9$ model at $T=0.55\, T^\ast$ and $k_x = k_y =k^\ast$. \emph{Left panel:} Contours of the vev of the operator dual to $\phi$, $\left<O_\phi\right>$. \emph{Right panel:} The charge density of the boundary field theory, with some integral curves of $\left<J^i\right>$ overlayed. We have shown four times the area of the computational domain at $N=40$.\label{prelimcharge}}
\end{center}
\end{figure}
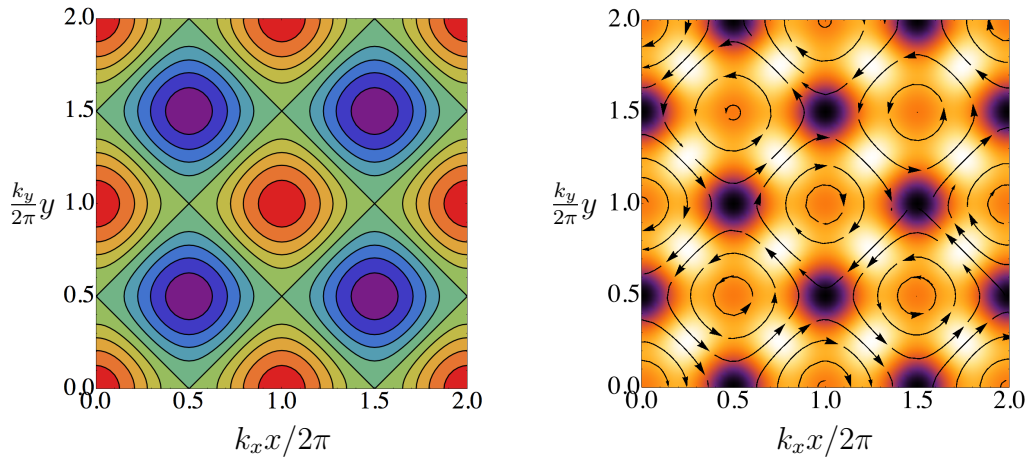

The scalar vev $\left<O_\phi\right>$ and charge density distribution is presented in Figure \ref{prelimcharge}, together with integral curves of the (divergence free) boundary current, $\left<J_i\right>$. The current is seen to circulate with the sense alternating between adjacent plaquettes of the checkerboard. This should not be too surprising; this would be qualitatively the picture formed for the current near $T^*$ simply by superposing two of the linear modes \eqref{instabnodef}. We emphasise however that this solution is not near the critical temperature, and is backreacted in the nonlinear regime. The convergence of $\xi$ and $\psi$ with the number of grid points for this solution shown in Appendix  \ref{convergence}, exhibiting exponentially fast convergence.

\subsection{Varying k\label{varyk}}
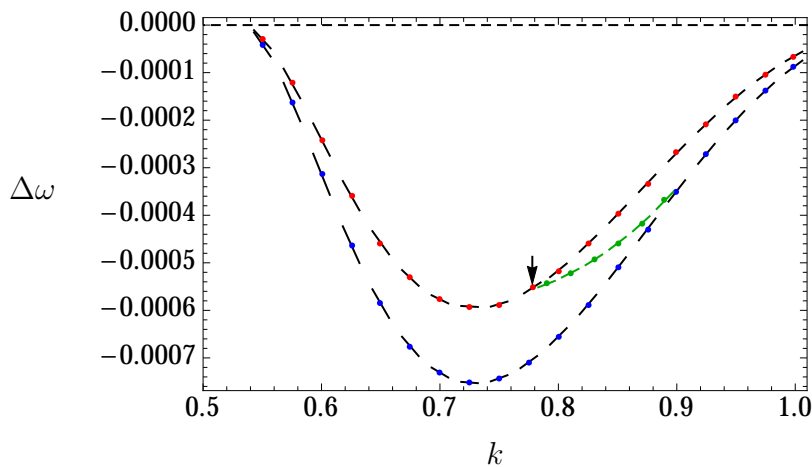
\begin{figure}[h!]
\begin{center}
\includegraphics[width=0.6\columnwidth]{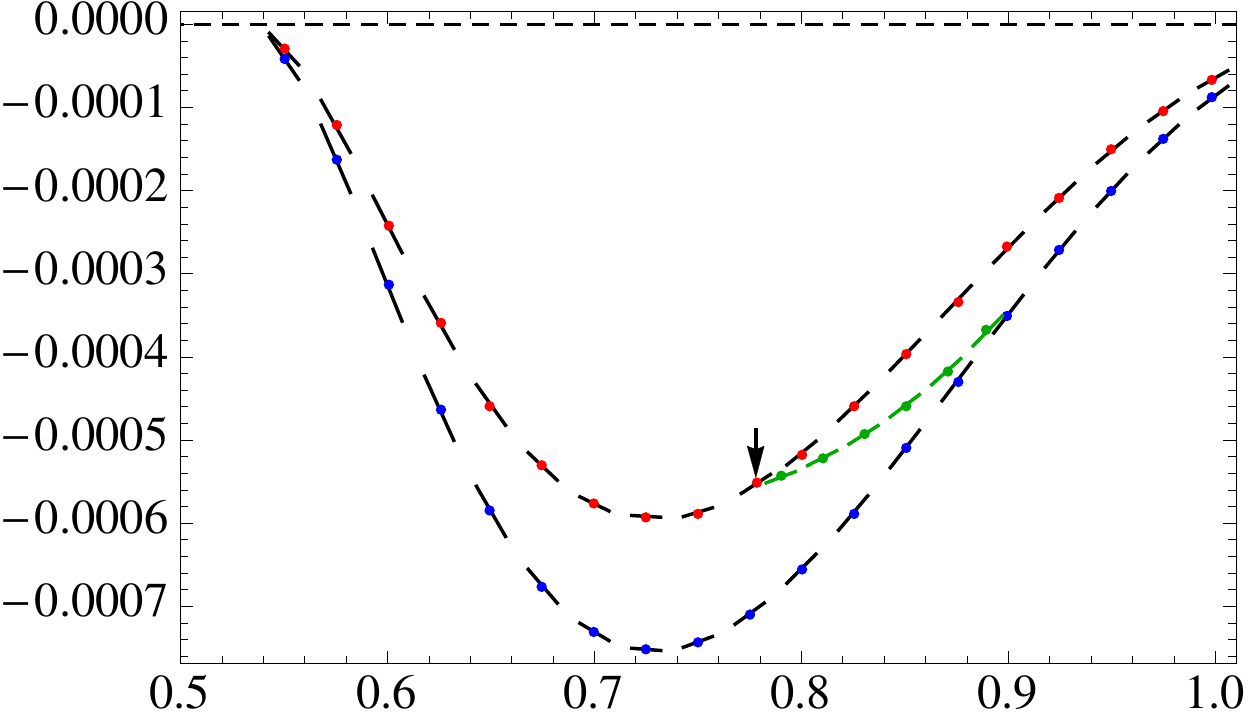}
\begin{picture}(0.1,0.1)(0,0)
\put(-300,90){\makebox(0,0){$\Delta \omega$}}
\put(-125,-10){\makebox(0,0){$k$}}
\end{picture}
\vskip 1em
\caption{Free energy difference with the normal phase, $\Delta \omega$, at fixed $T=0.55T^\ast$ in the $n=36,c_1 = 9.9$ model. The red points are square checkerboards with $k=k_x=k_y$, shown in conjunction with the black tangents computed using the first law, \ref{dwdk}, showing good agreement. The arrow marks the solution at the $k_x=k_y=k^\ast$ fiducial momentum scale. The blue points and tangents are the corresponding data for the striped phases with $k=k_y$. The remaining data in green are for non-square checkerboards, with fixed $k_x=k^\ast$ and varying $k=k_y> k^\ast$. \label{kscan}}
\end{center}
\end{figure}

At fixed $T$ there is a 2-parameter family of solutions parameterised by $k_x,k_y$. Both checkerboards and stripes exist within this family and are continuously connected, as we shall demonstrate in this section. We wish to minimise the free energy in this family. Defining, 
\be
\Delta \omega = \bar\omega - \omega_{normal} \label{deltaomega}
\ee
where $\omega_{normal}$ denotes $\bar\omega$ in the normal phase. $\Delta \omega$ for striped and `square' checkerboard solutions with $k\equiv k_x = k_y $ at fixed $T=0.55T^\ast$ and varying $k$ are shown in Figure \ref{kscan}. For this model we can see that the striped solutions are thermodynamically dominant, at least to the classes of checkerboards we have investigated in this section, where the dominant stripe has $\Delta \omega \simeq -7.5 \times 10^{-4}$ at $k\simeq 0.73$ (more details can be found in \cite{Withers:2013kva}). In the section \ref{transition} we will present a different set of model parameters where checkerboards are the dominant phase.

Following \cite{Donos:2013cka} (see also \cite{Rozali:2013ama}) we may compute the variation of the $\bar\omega$ with respect to the momenta. We find, 
\be
k_x\frac{\partial\bar{\omega}}{\partial k_x} = \bar\omega + \bar{T}^{x}_{~x} \label{dwdk}
\ee
and similarly for $y$. Note that if we look at the `square' solutions where $k_x=k_y$ we have by symmetry $k_x\frac{\partial\bar{\omega}}{\partial k_x} = k_y\frac{\partial\bar{\omega}}{\partial k_y}$. If we then use \eqref{dwdk} in an iterative scheme such as Newton-Raphson in order to find the minimum $\bar\omega$ starting with a $k_x=k_y$ checkerboard, we will stay  within the $k_x=k_y$ family. In practise we do not implement $k_x\frac{\partial\bar{\omega}}{\partial k_x} =0$ numerically e.g. as a boundary condition, but we do utilise the relation \eqref{dwdk} both in order to check our results and to assist in the determination of the minimum $\bar\omega$. We illustrate the agreement of the relation \eqref{dwdk} with the solutions constructed for the case of $k_x=k_y$ in Figure \ref{kscan}; the gradients computed in \eqref{dwdk} shown as tangents to the data, match the gradients of the data itself.

\begin{figure}[h!]
\begin{center}
\vspace{10pt}
\includegraphics[width=0.9\columnwidth]{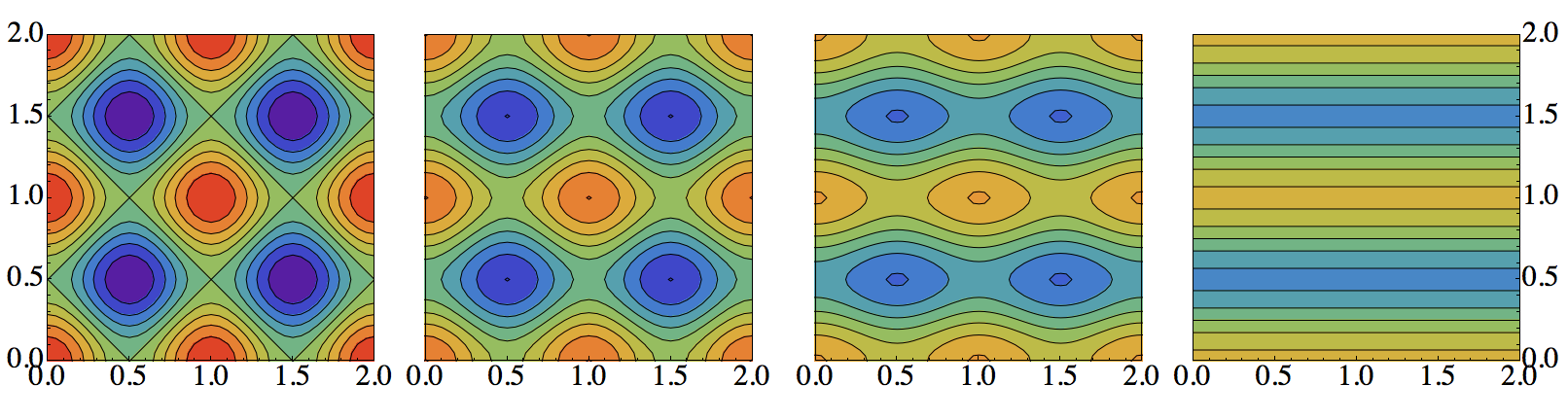}
\begin{picture}(0.1,0.1)(0,0)
\put(-420,60){\makebox(0,0){$\frac{k_y }{2\pi} y$}}
\put(-205,-10){\makebox(0,0){$k_x x / 2\pi$}}
\put(-353,110){\makebox(0,0){\footnotesize{$k_y=k^\ast$}}}
\put(-258,110){\makebox(0,0){\footnotesize{$k_y=0.87$}}}
\put(-155,110){\makebox(0,0){\footnotesize{$k_y=0.91$}}}
\put(-60,110){\makebox(0,0){\footnotesize{$k_y=0.925$}}}
\end{picture}
\vspace{10pt}
\caption{Continuously connecting checkerboards to stripes. We show contours of $\left<O_\phi\right>$ with wavenumbers $k_y = k^\ast, 0.87,0.91,0.925$ from left to right, keeping $k_x = k^\ast$ and $T=0.55 T^\ast$ fixed for the $n=36,c_1 =9.9$ model.\label{squash}}
\end{center}
\end{figure}
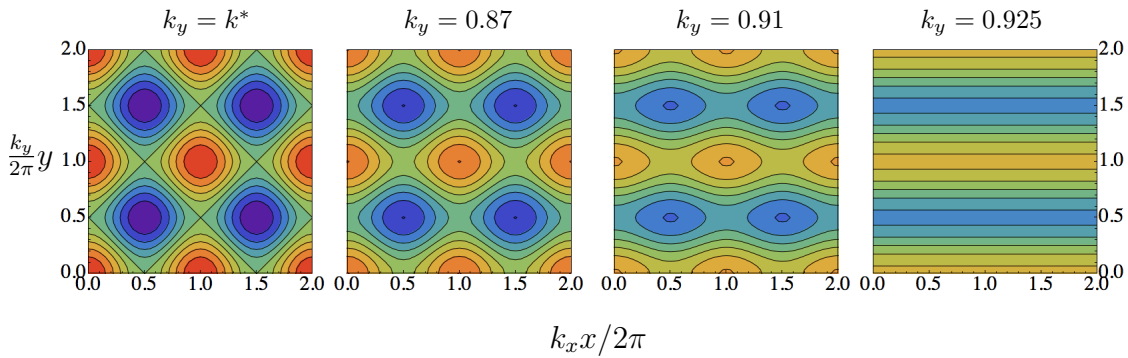

We would also like to understand how the checkerboard solutions presented in Figure \ref{prelimcharge} are related to the striped phases. In fact, they are continuously connected with striped solutions at the same temperature through the variation of $k_x$ and $k_y$. This is demonstrated by `squashing' the checkerboard in Figure \ref{squash}, where we keep $k_x = k^\ast$ fixed and dial $k_y$ at fixed $T=0.55 T^\ast$.
The free energy corresponding to this squashed branch of checkerboards is shown as the green points in Figure \ref{kscan}, joining the dominant square checkerboard branch with the striped branch.

\section{Modulated instabilities of $\phi\neq 0$ black branes\label{p1def}}
In this section we seek linear, marginal modes which indicate spatially modulated instabilities of black branes of the model \eqref{action} for which $\phi \neq 0$ in the normal phase at general $\tau,V,\vartheta$. For the examples of this paper, a black brane of this type will occur if $\phi_{(1)}\neq 0$, replacing RN as the normal phase of the system. Indeed we find these solutions are unstable, continuing the RN instabilities. In section \ref{transition} we construct the corresponding backreacted stripe and checkerboard solutions. 

The normal phase may be constructed by numerically solving a set of ODEs. We seek solutions of the form,
\ba
ds^2  = g_{MN}dx^M dx^N &=& \frac{1}{z^2} \left(-f(z) T(z) dt^2 + \frac{Z(z)}{f(z)}dz^2 + dx^2 + dy^2\right),\\
A(z,x^i) &=& a(z) dt,\qquad \phi(z,x^i) = \Phi(z)
\ea
where $f(z)$ is defined in section \ref{ansatzsec}. This results in a system of second order ODEs for matter fields $a,\Phi$ with first order equations for the metric functions $T,Z$. The construction proceeds via a standard shooting problem to enforce horizon regularity and boundary normalisability, see for example \cite{Gauntlett:2009bh,Donos:2012yu}. Counting the number of undetermined coefficients in the near horizon expansion (there are 3 with the horizon position fixed at $z=1$) and near boundary expansion (5) reveals solutions will exist in two-parameter families, as expected. We may take convenient parameters to be the source for $O_\phi$ (this is $\phi_1/\mu$ for the $m^2 = -2$ case) and the temperature $T/\mu$. The zero temperature solutions will be discussed for specific examples later in this section. 

The spatially modulated instability of the RN solutions involving the fluctuations \eqref{instabnodef} continues to the $\phi\neq 0$ branes. However,  because the normal phase has $\Phi(z)\neq 0$  \eqref{instabnodef} no longer forms a consistent set of perturbations. We find it convenient to work in a particular gauge with $\delta g_{tt}=0$ and $\delta g_{zx}=0$, which fixes a set of gauge modes arising from first-order coordinate transformations of the background solution. Here a consistent set of perturbations for any $\tau,V,\vartheta$ consists of second order ODEs for \eqref{instabnodef} together with
\ba
\delta A_t(z,x) &=&  a_{t}(z) \cos{k x} \label{p1instab1}\\
\delta g_{ii}(z,x)&=& h_{ii}(z) \cos{k x}\qquad (i=1,2)\\
\delta g_{zz}(z,x) &=& h_{zz}(z) \cos{k x}. \label{p1instab4}
\ea
We may write the equations as second order for $a_{t}(z)$, and first order for $h_{xx}(z), h_{yy}(z)$ and $h_{zz}(z)$. 

Note that due to $a_t$, the charge density may become modulated at leading order with a momentum $k$ rather than at sub-leading order with momentum $2k$ as for the instabilities of the normal phase with $\phi=0$. This leading order modulation can be seen explicitly in the nonlinear checkerboard and stripe solutions constructed in section \ref{transition}, with the charge density modulated with the same period as the scalar field $\phi$, both near the critical temperature and also extending to solutions with significant backreaction. 

The fluctuations comprise a coupled system of ODEs with total differential order 11. At the horizon the fields take the form, 
\ba
\lambda(z) &=& \lambda^{(0)} +O(1-z) \\
a_{t}(z) &=& a_{t}^{(1)}(1-z) +O(1-z)^2 \\
a_{y}(z) &=& a_{y}^{(0)} +O(1-z)\\
h_{ii}(z) &=& h_{ii}^{(0)} + O(1-z)\\
h_{ty}(z) &=& h_{ty}^{(1)}(1-z) + O(1-z)^2\\
h_{zz}(z) &=& O(1)
\ea
where we have shown any undetermined coefficients in the near-horizon expansion. In the UV we enforce normalisability, 
\ba
\lambda(z) &=& \lambda^{(2)}z^2 +O(z)^3 \\
a_{t}(z) &=& a_{t}^{(1)}z +O(z)^2 \\
a_{y}(z) &=& a_{y}^{(1)}z +O(z)^2\\
h_{ii}(z) &=& O(z)\\
h_{ty}(z) &=& h_{ty}^{(3)}z + O(z)^2\\
h_{zz}(z) &=& h_{zz}^{(3)}z + O(z)^2
\ea
in particular note that none of the boundary metric components, $\mu$ or the source $\phi_{(1)}$ are affected by this mode, and remain constant as required for spontaneous modulation. There are 11 undetermined coefficients at linear order, one of which may be fixed by linearity. Coupled together with the equations for the normal phase, the total differential order is 17.  To match this, there are 18 undetermined coefficients and the value of the momentum $k$. Thus we expect 2-parameter families of critical solutions, which we can parameterise by $k/\mu$ and $\phi_{(1)}/\mu$. The $\phi \neq 0$ normal phase and its instabilities for the model class \eqref{taudef} in the case $n=0$ is presented in section \ref{instab0}, in preparation for the backreacted solutions of this model, studied in section \ref{transition}.

\subsection{$n=0$\label{instab0}}
\begin{figure}[h!]
\begin{center}
\includegraphics[width=0.55\columnwidth]{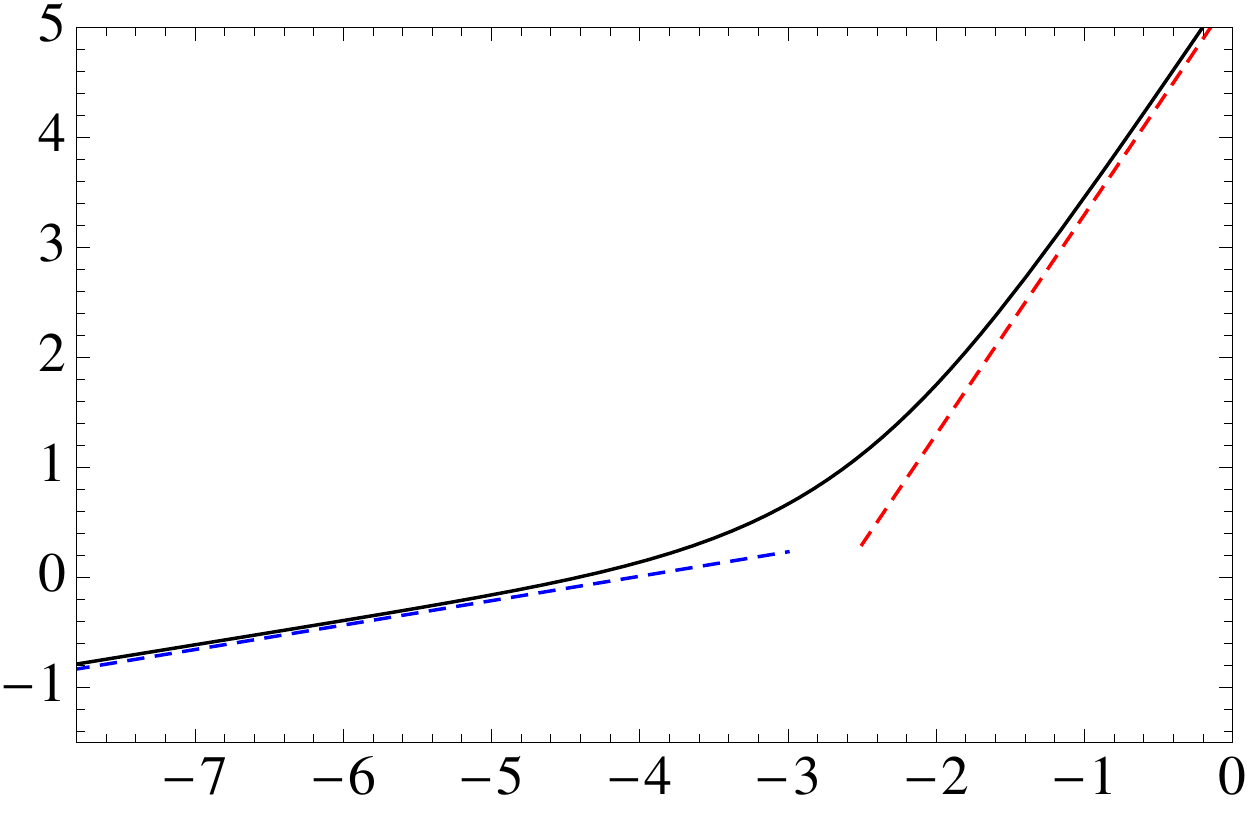}
\begin{picture}(0.1,0.1)(0,0)
\put(-270,100){\makebox(0,0){$\log{s}$}}
\put(-170,30){\makebox(0,0){$s\propto T^{2/9}$}}
\put(-120,-10){\makebox(0,0){$\log{T}$}}
\put(-35,80){\makebox(0,0){$s\propto T^{2}$}}
\end{picture}
\vskip 1em
\caption{The scaling of the entropy density, $s$, with temperature for normal phase in the $n=0$ model at $\phi_{(1)} = 0.5$. At high temperatures the behaviour is $s\propto T^2$, indicated by the dashed red line. At low temperatures the dashed blue line shows the behaviour $s\propto T^{2/9}$ consistent with the emergence of the hyperscaling-violating geometry \eqref{hsv1}-\eqref{hsv3} in the IR at low temperatures. \label{n0entropy}}
\end{center}
\end{figure}
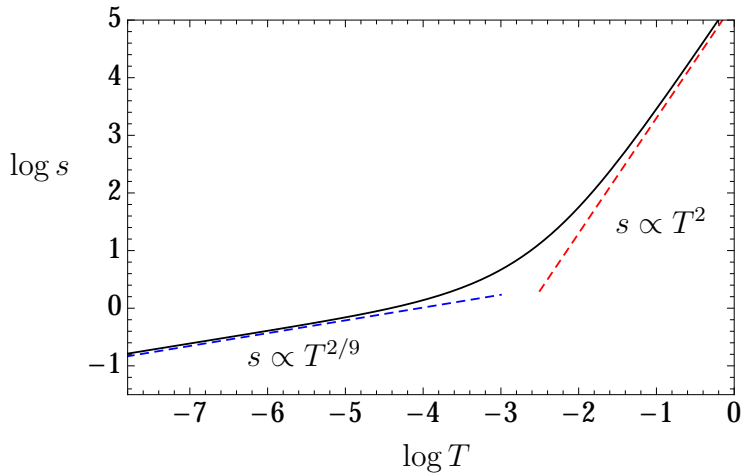

When $n=0$ it is known that the AdS$_2\times R^2$ near-horizon geometry RN is linearly unstable to $k=0$ modes involving the scalar \cite{Denef:2009tp, Donos:2011bh}. Thus we may expect that turning on $\phi_{(1)}$ can drive the IR away from the $\phi=0$ AdS$_2\times R^2$ to something else, such as a hyperscaling violating (HSV) geometry with a running scalar. Indeed, the theory admits the HSV solutions \cite{Charmousis:2010zz,Gouteraux:2011ce},
\ba
ds^2 &=& -z^{11/2}dt^2+\frac{11e^{\frac{\sigma}{\sqrt{3}}}}{4}\frac{dz^2}{z} +\sqrt{z}\left(dx^2+dy^2\right)\label{hsv1}\\
\phi &=& \sqrt{3}\log z + \sigma\label{hsv2}\\
A &=& 2\sqrt{\frac{5}{11}} z^{11/4}dt\label{hsv3}
\ea
This solution has a hyperscaling violation exponent $\theta=4$ and dynamical critical exponent $z=-9$. Based on this scaling property we can infer the low temperature scaling of the entropy density, $s\propto T^{2/9}$. We confirm this expectation by numerically constructing the finite temperature branch, for which the entropy density is shown in Figure \ref{n0entropy}. 

\begin{figure}[h!]
\begin{center}
\includegraphics[width=0.5\columnwidth]{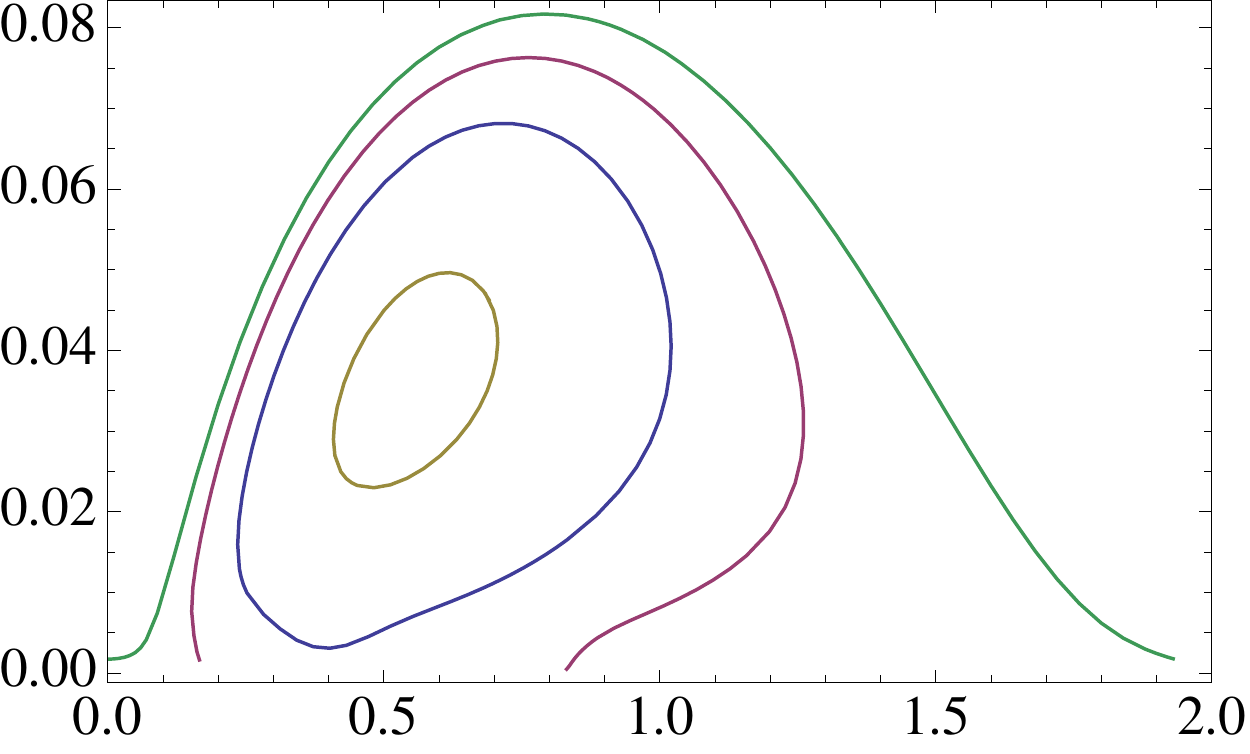}
\begin{picture}(0.1,0.1)(0,0)
\put(-245,80){\makebox(0,0){$T$}}
\put(-110,-10){\makebox(0,0){$k$}}
\put(-40,45){\makebox(0,0){\scriptsize{$0$}}}
\put(-75,50){\makebox(0,0){\scriptsize{$0.08$}}}
\put(-100,60){\makebox(0,0){\scriptsize{$0.12$}}}
\put(-130,70){\makebox(0,0){\scriptsize{$0.16$}}}
\end{picture}
\vskip 1em
\caption{Critical temperatures for spatially modulated linear instabilities of the $\phi_{(1)}$-deformed normal phase solutions in the case $n=0$, as a function of the wavenumber $k$ of the modulation. The labels shown are the values of $\phi_{(1)}/\mu$ which is fixed for each curve.
 \label{belln0}}
\end{center}
\end{figure}
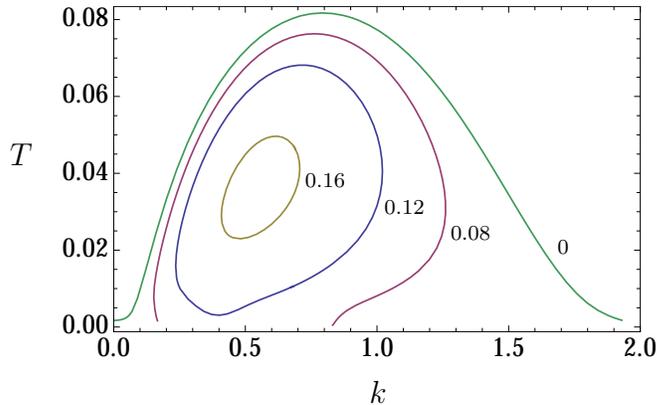

Hence we see that dialling the parameter $\phi_{(1)}$ results in at least one quantum phase transition in the normal phase, moving from the $\phi = 0$ AdS$_2\times R^2$ to the HSV geometry \eqref{hsv1}-\eqref{hsv3}. It is interesting to note the effect that this change of IR has on the spatially modulated instabilities \eqref{p1instab1}-\eqref{p1instab4}. Since the AdS$_2\times R^2$ case is $k\neq 0$ unstable \cite{Donos:2011bh} by continuity we expect that the instability survives for small $\phi_{(1)}\neq 0$, at least for finite temperature. Indeed, the critical temperature bell-curves do continue to $\phi_{(1)}\neq 0$ as shown in Figure \ref{belln0}. 
The change is most striking at lower temperatures where the linear instability at $T=0$ seems to disappear entirely, at least for large enough $\phi_{(1)}$. It would be interesting to investigate this feature directly, and any connection with $k\neq 0$ stability properties of the HSV infrared geometry.

\section{Checkerboard transitions \label{transition}}
The checkerboards of section \ref{prelim} are thermodynamically subdominant to the striped solutions. In this section we explore the role of introducing a homogeneous source $\phi_{(1)}$ for the operator dual to the scalar $\phi$. Linear instabilities in this case were constructed in section \ref{p1def}. Here we consider the case $n=0$ with associated instabilities mapped out in section \ref{instab0}. The key results of this section are checkerboards which are thermodynamically dominant to the stripes, including first order normal-to-checkerboard first order transitions followed by checkerboard-to-stripe first order transitions at lower temperatures.

\begin{figure}[h!]
\begin{center}
\includegraphics[width=0.35\columnwidth]{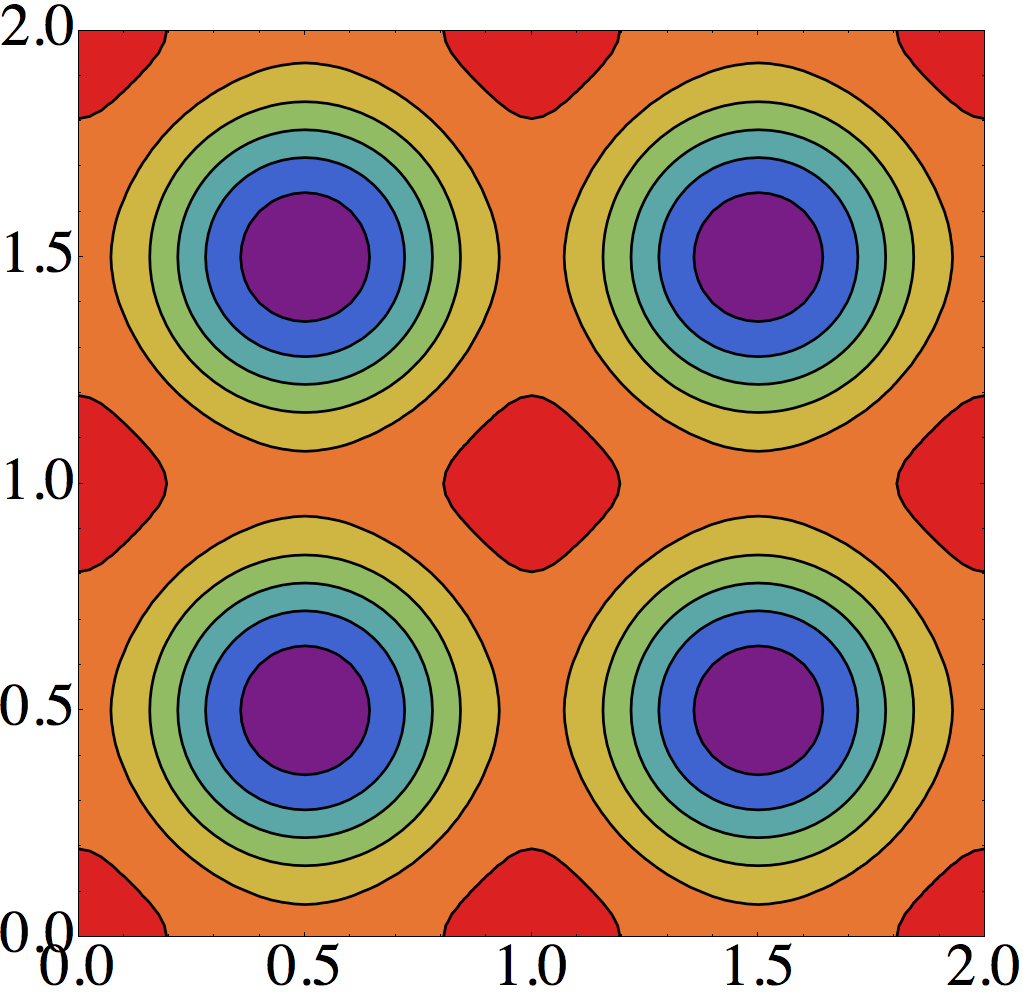}
\hspace{40pt}
\includegraphics[width=0.35\columnwidth]{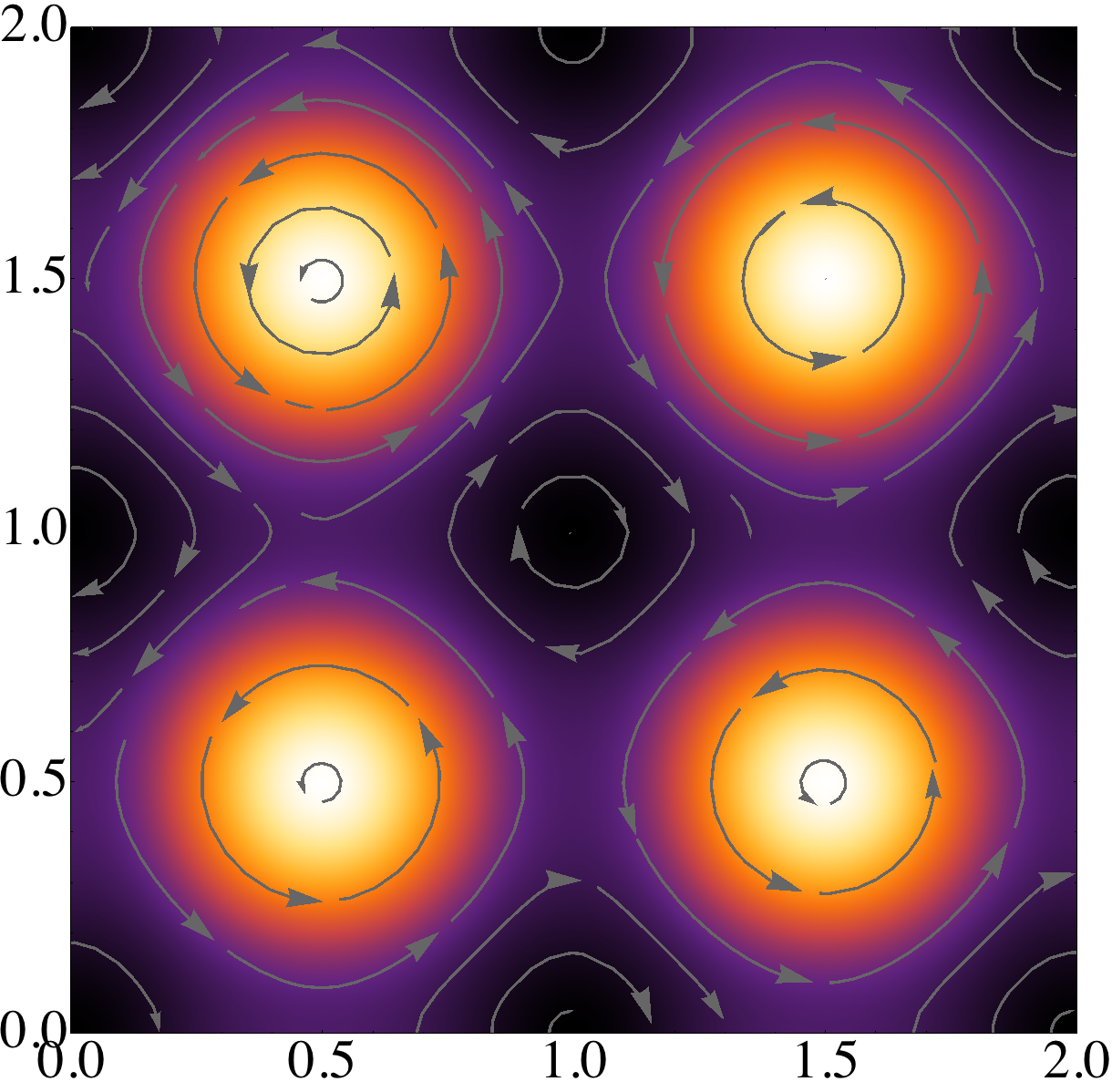}
\begin{picture}(0.1,0.1)(0,0)
\put(-380,80){\makebox(0,0){$\frac{k_y }{2\pi} y$}}
\put(-80,-10){\makebox(0,0){$k_x x / 2\pi$}}
\put(-175,80){\makebox(0,0){$\frac{k_y}{2\pi} y$}}
\put(-285,-10){\makebox(0,0){$k_x x / 2\pi$}}
\end{picture}
\vskip 1em
\caption{
A holographic checkerboard in the $n=0,c_1=9.9$ model at $T=0.57\, T_c$ near the preferred $k_x = k_y$ value for a constant deformation $\phi_{(1)} = 0.13$. \emph{Left panel:} A contour plot of the vev of the operator dual to $\phi$, $\left<O_\phi\right>$. \emph{Right panel:} The charge density of the boundary field theory, with some integral curves of $\left<J^i\right>$ overlayed. \label{charge13}}
\end{center}
\end{figure}
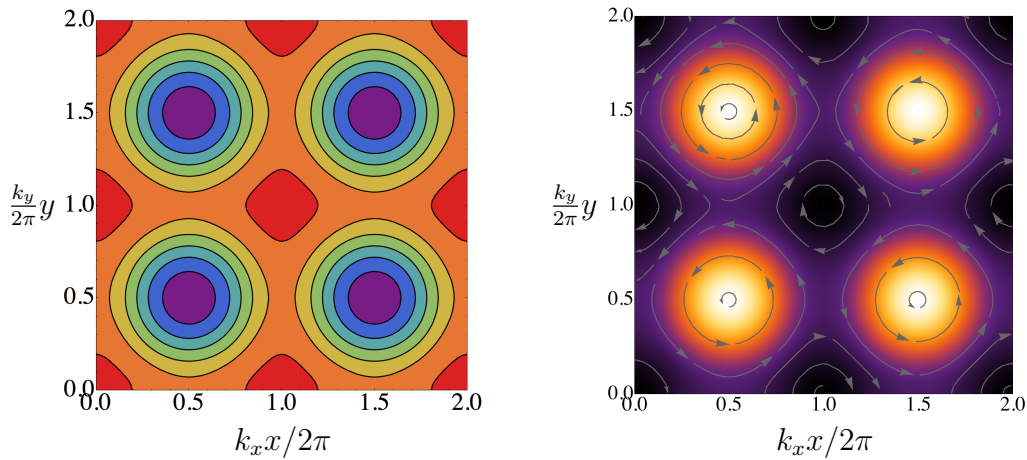

An example of a checkerboard in the $\phi_{(1)}$-deformed model is shown in Figure \ref{charge13}, at $T=0.57T_c$ where $T_c\simeq 0.0707$  is the critical temperature of the normal-to-checkerboard phase transition demonstrated shortly.  We show convergence tests of the numerics for this solution, presented in Appendix \ref{convergence}. 

\begin{figure}[h!]
\begin{center}
\includegraphics[width=0.48\columnwidth]{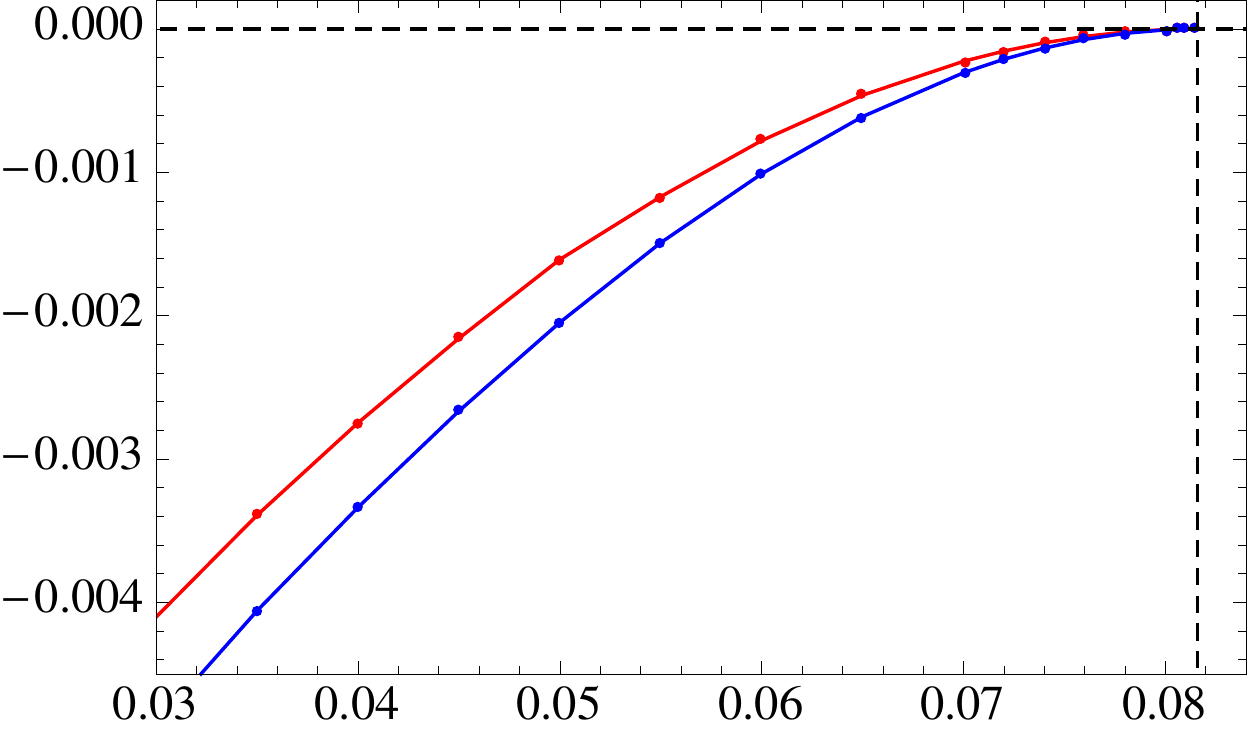}
\begin{picture}(0.1,0.1)(0,0)
\put(-230,80){\makebox(0,0){$\Delta \omega$}}
\put(-110,-10){\makebox(0,0){$T$}}
\put(-160,115){\makebox(0,0){\footnotesize$\phi_{(1)} = 0$}}
\end{picture}
\vskip 1em
\caption{Free energy difference of the square checkerboard (red) and striped (blue) solutions with the normal phase, for the model $c_1=9.9,n=0$ with $\phi_{(1)} = 0$. Each checkerboard point is obtained by minimising $\bar\omega$ with respect to $k_x=k_y$, and similarly for the stripe solutions. The vertical dashed line indicates the transition temperature and the linear instability temperature, $T\simeq 0.0816$, computed using the analysis of section \ref{p1def}.\label{freenodef}}
\end{center}
\end{figure}
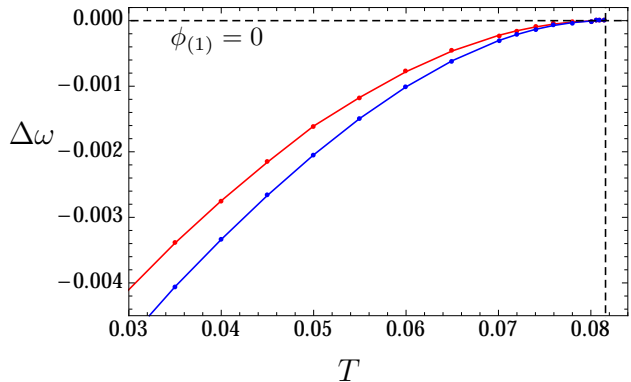

At each temperature we construct the thermodynamically dominant solution of the square checkerboard family by minimising $\Delta \omega$ with respect to $k_x= k_y$, as described in section \ref{varyk}. Similarly we construct the dominant stripe solutions. The temperature dependence for the $\phi_{(1)} = 0$ case is shown in Figure \ref{freenodef}. Clearly the striped solutions are thermodynamically dominant, as in the $n=36$ case discussed in section \ref{prelim}.

\begin{figure}[h!]
\begin{center}
\includegraphics[height=0.25\columnwidth]{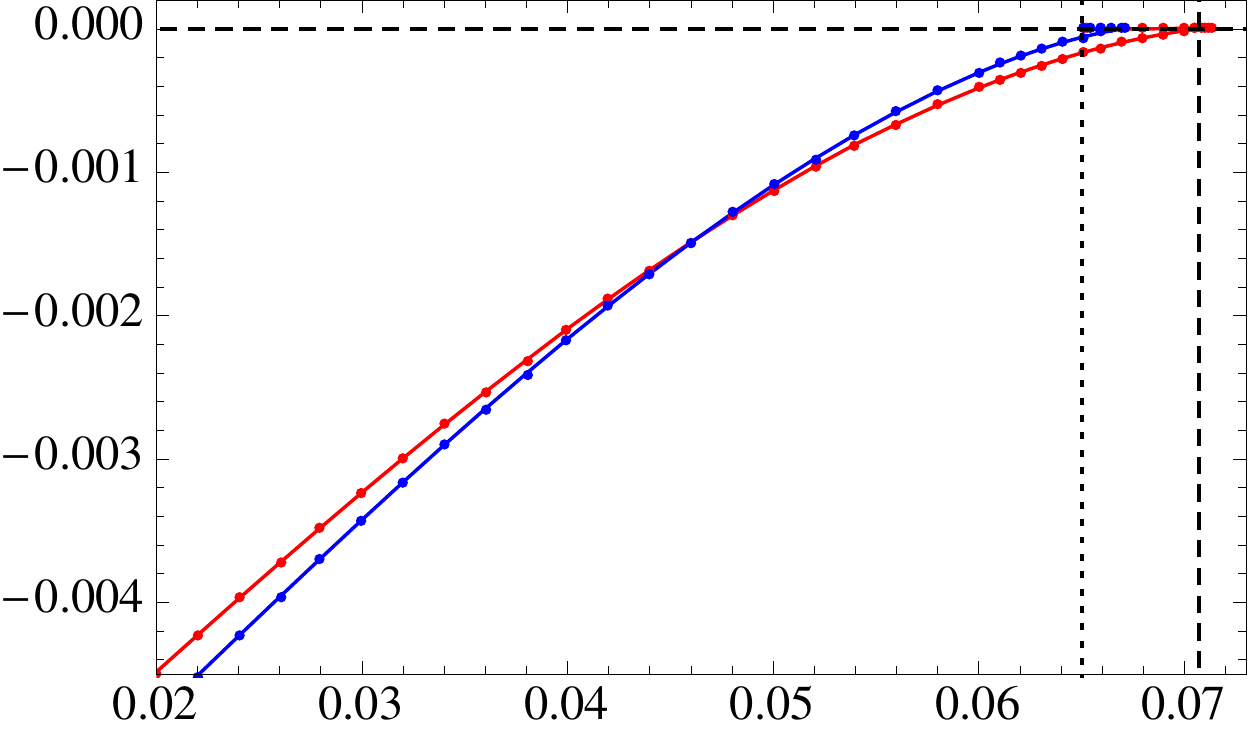}
\hspace{30pt}
\includegraphics[height=0.25\columnwidth]{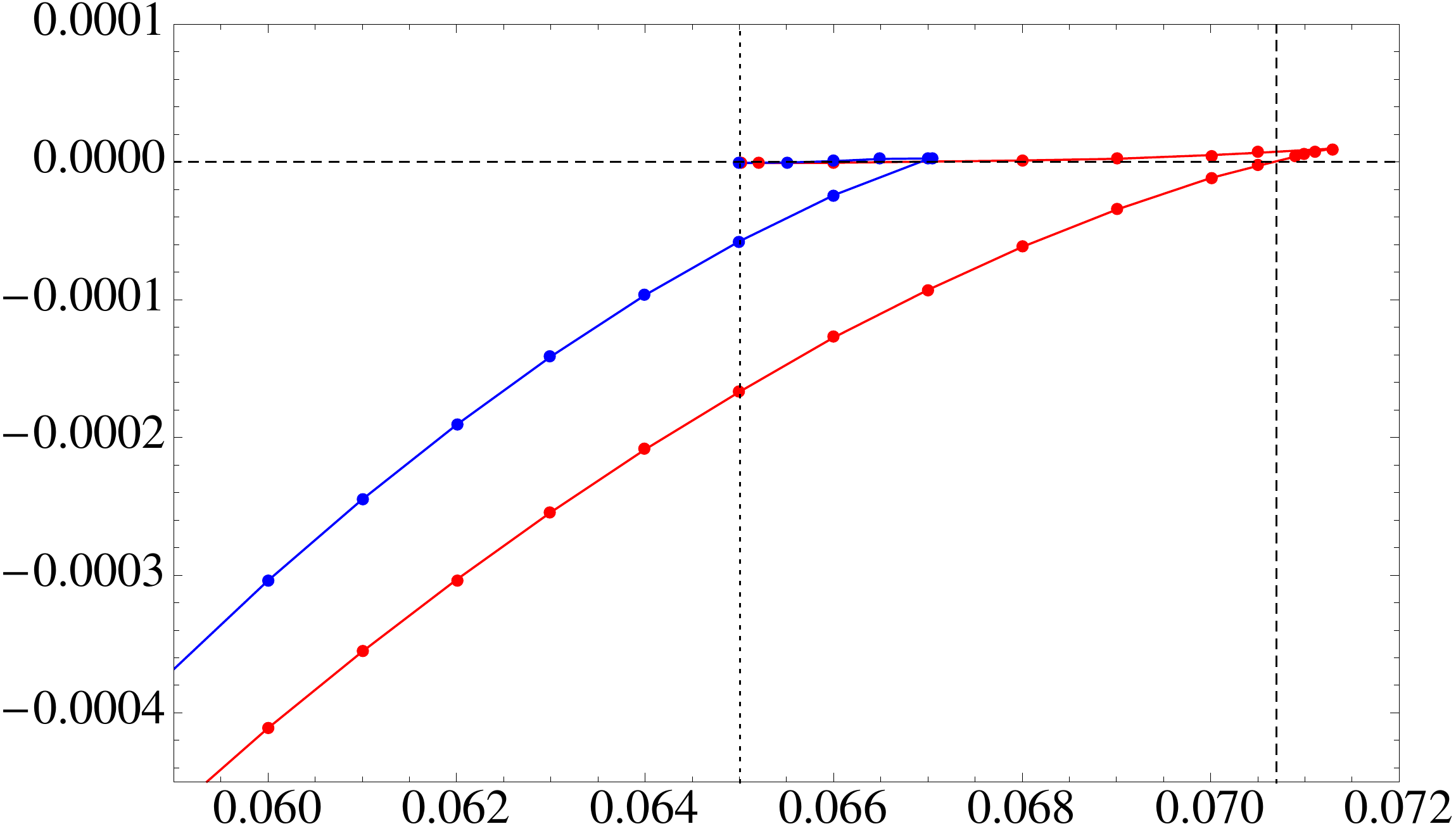}
\begin{picture}(0.1,0.1)(0,0)
\put(-210,80){\makebox(0,0){$\Delta \omega$}}
\put(-90,-10){\makebox(0,0){$T$}}
\put(-430,75){\makebox(0,0){$\Delta \omega$}}
\put(-330,-10){\makebox(0,0){$T$}}
\put(-370,100){\makebox(0,0){\footnotesize$\phi_{(1)} = 0.13$}}
\put(-140,100){\makebox(0,0){\footnotesize$\phi_{(1)} = 0.13$}}
\end{picture}
\vskip 1em
\caption{Free energy difference of the square checkerboard (red) and striped (blue) solutions with the normal phase, for the model $c_1=9.9,n=0$ with $\phi_{(1)} = 0.13$. Each checkerboard point is obtained by minimising $\bar\omega$ with respect to $k_x=k_y$, and similarly for the stripe solutions. The right panel shows the swallowtail region and the first order normal-to-checkerboard phase transition in more detail. On the left panel the first order checkerboard-to-stripe phase transition can be seen at lower temperatures. The vertical dotted line is the marginal mode temperature, $T\simeq 0.0650$, computed directly in the linear analysis of section \ref{p1def}. The vertical dashed line shows the position of the checkerboard transition, $T\simeq 0.0707$. \label{freedef}}
\end{center}
\end{figure}
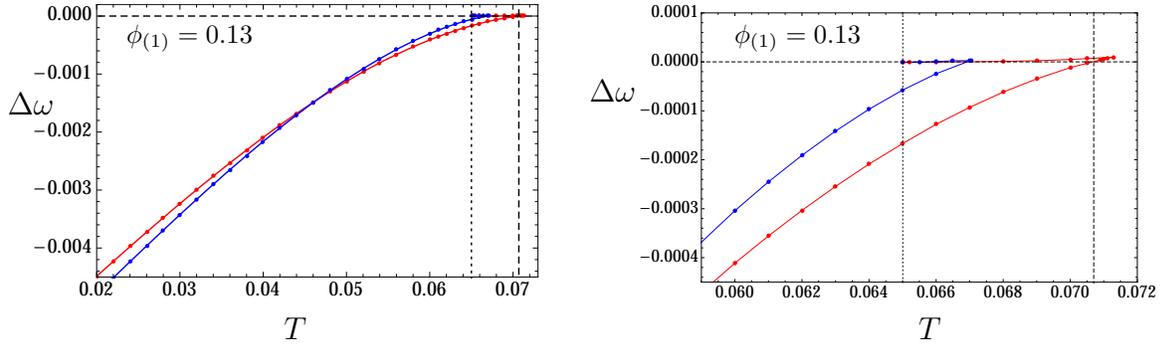

However the crucial behaviour of this model lies in the $\phi^{(1)}$ deformations. In Figure \ref{freedef} we show the temperature dependence of $\Delta \omega$ for the case $\phi_{(1)} = 0.13$. The introduction of $\phi^{(1)}$ has opened a swallow-tail structure near the instability threshold, pushing the checkerboard transition temperature higher than that of the stripes. This results in a first order phase transition from the normal phase to the checkerboard phase, followed by another first order phase transition to the striped phase at lower temperatures.

\begin{figure}[h!]
\begin{center}
\begin{tikzpicture}[scale=0.6]
	\draw[->] (-0.2,0) -- (9.5,0) node[right] {$\phi_{(1)}$};
	\draw[->] (0,-0.2) -- (0,6.4) node[above] {$T$};
   	\def\triple{(4,4)};
   	\clip[clip] (-0.1,-0.1) rectangle (9.5,6.4);
   	\def\stop{ (0,6) .. controls (2.4,5.5)  .. \triple};
  	\def\ctop{ \triple .. controls (7.2,3.2)  .. (9.4,0.5)};
   	\def\firsto{ \triple .. controls (4.32,3.0)  .. (5.2,0.5)};
     	\def\firstor{ (4.8,0) .. controls (4.32,3.0)  .. \triple};
	\draw[blue!0!black, line width=1] \stop ;
	\draw[blue!0!black, line width=1,dashed] \ctop ;
	\draw[blue!0!black, line width=1,dotted] \firsto ;
	\draw (1.8,2) node[] {Stripes};
	\draw (6.55,2) node[below] {Checker-};
	\draw (6.9,1.4) node[below] {boards};
	\draw (5.8,5.3) node[above] {Normal};
	\draw[black] \triple circle(1mm);
	\fill[fill=black] \triple circle(1mm);
\end{tikzpicture}
\caption{A schematic $T-\phi_{(1)}$ phase diagram for the $c_1=9.9,n=0$ model, inferred from data obtained at several fixed-$\phi_{(1)}$ slices,  excluding the low temperature region. Solid lines denote second order phase transitions whilst the dashed and dotted lines indicate first order phase transitions, with a tri-critical point labelled by the dot.\label{cartoonphase}}
\end{center}
\end{figure}
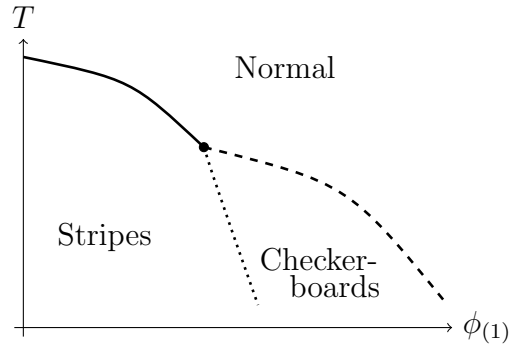

Considering several other fixed-$\phi_{(1)}$ slices, we find that the temperature of the checkerboard to stripe first-order phase transition decreases as $\phi_{(1)}$ is increased. At sufficiently small, positive $\phi_{(1)}$ there is no checkerboard transition at all. At higher values of $\phi_{(1)}$ the checkerboard phase is dominant for all temperatures considered, although note we have no data at very low temperatures. Based on these observations we anticipate the qualitative structure of the phase diagram illustrated in Figure \ref{cartoonphase}. Note that the tri-critical point occurs at a value of $\phi_{(1)}$ lower than that required for the linear instabilities of section \ref{instab0} to disappear. 

\section{Discussion \label{comments}}

We have constructed cohomogeneity-three, finite temperature stationary black brane solutions dual to a CFT exhibiting checkerboard order.  Due to the parity violating term \eqref{action} the phase breaks $P$ and $T$ resulting in $\left<J_i\right>\neq 0$ which circulates within each checkerboard plaquette. The phase appears spontaneously, without any UV lattice or other inhomogeneities introduced by hand.

We found qualitatively different behaviour depending on the model parameters, $(n,c_1)$ as well as the constant deformation governed by the source $\phi_{(1)}$.
The checkerboards constructed are thermodynamically preferred over the normal phase below a critical temperature, with a first order transition depending on the model.
We also explored competition with the striped phases.
At $n=0,c_1=9.9$ and $\phi_{(1)}\neq 0$ we found that there can be up to two first-order phase transitions, from the normal phase to the checkerboard phase and from the checkerboard phase to the striped phase at lower temperatures. 

We constructed linear $k\neq 0$ instabilities of normal phase solutions with $\phi \neq 0$. Here, the charge density becomes modulated at leading order in the instability at wavenumber $k$. We investigated one model where the IR becomes hyperscaling-violating as the deformation parameter $\phi_{(1)}$ is increased. It would be interesting to extend this analysis and investigate $T=0$ $k\neq 0$ stability directly in the absence of an AdS$_2\times R^2$ IR solution. 

A remaining open question is the nature of the $T=0$ ground states of both the striped and checkerboard phases. It would be particularly interesting to see the explicit $T=0$ manifestation of the checkerboard to stripe phase transition in the phase diagram of Figure \ref{cartoonphase}. Our numerical analysis has restricted to the case where the checkerboards are rectangular. There may be a larger space of solutions and it would be interesting to investigate the dominant shape of these phases.\footnote{See \cite{Bu:2012mq} for an interesting study of dominant lattice shapes for a magnetic induced instability in the probe limit.}

Finally we comment on recent progress in the construction of \emph{explicitly} modulated phases utilising spatially dependent field theory sources and bulk fields in such a way that only ODEs need to be solved in the bulk \cite{Donos:2013eha, Andrade:2013gsa}\footnote{See related work in the holographic massive gravity literature \cite{Vegh:2013sk, Davison:2013jba,Blake:2013owa}.}. Note that the approach taken in \cite{Andrade:2013gsa} allows for analytic construction of the background geometry\footnote{The solutions were previously constructed in \cite{Bardoux:2012aw} and are the same as those seen in the massive gravity context. See also \cite{Taylor:2014tka} for analytic solutions with with non-canonical kinetic terms.} though note it may be generalised to other contexts by adding a dilaton, as explored in \cite{Gouteraux:2014hca}. These are technically similar in spirit to earlier work in 5d where helical symmetry is exploited to reduce a problem to ODEs \cite{Nakamura:2009tf,Iizuka:2012iv,Donos:2012gg}. It is conceivable that the mechanisms of \cite{Donos:2013eha, Andrade:2013gsa} may be used in the study of spontaneous modulation.\footnote{Indeed, \cite{Nakamura:2009tf,Donos:2012gg} involve the spontaneous development of helical order.} One concern in the present context is that, since they are restricted to special scalar configurations which allow the trick to work, they may not be capable of capturing the dominant phase.

\section*{Acknowledgements}
I would like to thank Toby Wiseman for valuable discussions on numerical issues.
I would also like to thank Aristomenis Donos and Jerome Gauntlett for early discussions.
I acknowledge the use of the IRIDIS High Performance Computing Facility and associated support services at the University of Southampton, in the completion of this work.

\appendix
\section{Holographic renormalisation\label{apprenorm}}
First consider the components of the metric organised according to $z$-directions and all others, 
\be
ds^2 = g_{zz}(z,x) dz^2 + 2 g_{z\mu}(z,x) dz dx^\mu + g_{\mu\nu}(z,x) dx^\mu dx^\nu
\ee
The near boundary expansion to the order of interest takes the following form,
\ba
g_{\mu\nu} &=& \frac{1}{z^2} \left(g^{(0)}_{\mu\nu} + g^{(2)}_{\mu\nu}(x)  z^2 + g^{(3)}_{\mu\nu}(x)  z^3 + O(z)^4\right)\\
g_{z\mu} &=& z\, g^{(2)}_{z\mu}(x)  +O(z)\\
g_{zz} &=& \frac{1}{z^2} \left(1  + g^{(2)}_{zz}(x)  z^2 + g^{(3)}_{zz}(x)  z^3 + O(z)^4\right).
\ea
We now consider the computation of the one-point functions, performing the holographic renormalisation procedure in Fefferman-Graham coordinates near the boundary. In particular, near the boundary we expand in the following coordinate system, 
\be
ds^2 = \frac{dr^2}{r^2} +\gamma_{\mu\nu}(r,\tilde{x}) d\tilde{x}^\mu d\tilde{x}^\nu
\ee
where $r=0$ gives the boundary.

The relation between these two coordinate systems in the vicinity of the boundary may be expressed as,
\ba
z &=& r\left(1+ R^{(2)}(\tilde{x}) r^2+R^{(3)}(\tilde{x}) r^3 + O(r)^4\right)\label{rdef}\\ 
x^\mu &=& \tilde{x}^\mu + O(r)^4\label{xtildedef}
\ea
where the transformation is given by the following functions, 
\ba
R^{(2)}(\tilde{x})  &=& -\frac{1}{4}g^{(2)}_{zz}(\tilde{x})\\
R^{(3)}(\tilde{x})  &=& -\frac{1}{6}g^{(3)}_{zz}(\tilde{x}).
\ea
The asymptotic form of the metric $\gamma$ can now be written in terms of the components of the original metric using the transformation above, 
\be
\gamma_{\mu\nu} =\frac{1}{r^2}\left(g_{\mu\nu}^{(0)} + \left(g_{\mu\nu}^{(2)} + \frac{1}{2} g_{\mu\nu}^{(0)}g_{zz}^{(2)}\right)r^2+ \left(g_{\mu\nu}^{(3)} + \frac{1}{3} g_{\mu\nu}^{(0)}g_{zz}^{(3)} \right)r^3 + \ldots\right).
\ee
The near-boundary expansion of the scalar field becomes, 
\ba
\phi &=& \phi^{(1)}(x) z+ \phi^{(2)}(x) z^2 + O(z)^3\\
&=&  \phi^{(1)}(\tilde{x}) r+ \phi^{(2)}(\tilde{x}) r^2 + O(r)^3.
\ea
Similarly, we may re-write the asymptotic expansions for the gauge field, 
\ba
A &=& \left(z A_z^{(1)}(x)+O(z)^2\right) dz +  \left( A_\mu^{(0)}(x) +z A_\mu^{(1)}(x)+O(z)^2\right) dx^\mu\\
&=& \left(r A_z^{(1)}(\tilde{x})+O(r)^2\right) dr +  \left( A_\mu^{(0)}(\tilde{x}) +r A_\mu^{(1)}(\tilde{x})+O(r)^2\right) d\tilde{x}^\mu
\ea
where the cross terms appear too high in order $z,r$ to affect the expressions. We may also perform a gauge transformation to eliminate the $A_r$ component near the boundary,
\be
\tilde{A}_M = A_M - \partial_M \lambda(r, \tilde{x}) \quad \text{with} \quad \lambda = \frac{1}{2} r^2 A_z^{(1)}(\tilde{x})
\ee
which results in
\be
\tilde{A} = O(r)^2  dr +  \left( A_\mu^{(0)}(\tilde{x}) +r A_\mu^{(1)}(\tilde{x})+O(r)^2\right) d\tilde{x}^\mu.
\ee

\subsection{One point functions}
In this section we use the coordinates $(r,\tilde{x})$, defined in \eqref{rdef},\eqref{xtildedef}.
First we define the unit one-form
\be
n = N dr = -\frac{1}{r}dr.
\ee
This is normal to the boundary at $r=0$. We can then define the orthogonal projector, 
\be
p_{MN} = g_{MN} - n_M n_N
\ee
In particular, $p_{rA}=0$, whereas,
\be
p_{\mu\nu} = \gamma_{\mu\nu}.
\ee
The extrinsic curvature is then given by
\be
K_{AB} = - p_A^{~C}p_B^{~D}\nabla_{(C} n_{D)}
\ee
similarly, $K_{rA}=0$ and,
\be
K_{\mu\nu} = \frac{r}{2} \partial_r \gamma_{\mu\nu}.
\ee
Expanding for small $r$, we find, 
\be
K_{\mu\nu} = -\frac{1}{r^2} \gamma^{(0)}_{\mu\nu} + \frac{1}{2} r \gamma_{\mu\nu}^{(3)} + O(r)^2. \label{appkdef}
\ee
The renormalised action is given by,
\be
S = S_b - 2\int d^3x \sqrt{-\gamma} K + 2 S_{ct}
\ee
with associated stress tensor, 
\be
T^{\mu\nu} = 2\left(K^{\mu\nu} - K \gamma^{\mu\nu} + \frac{2}{\sqrt{-\gamma}} \frac{\delta S_{ct}}{\delta \gamma_{\mu\nu}}\right)\label{apptdef}
\ee
A standard analysis \cite{Balasubramanian:1999re, Bianchi:2001kw} reveals the counterterms,
\be
S_{ct} = \int d^3x \sqrt{-\gamma}\left(-2 - \frac{1}{4}\phi^2\right).
\ee
Here we focus on flat boundary metric and we have omitted curvature counterterms. Combining \eqref{appkdef} and \eqref{apptdef} and the on-shell relation,
\be
\gamma_{(2)}^{\mu\nu} -  \text{tr} \gamma^{(2)} \gamma_{(0)}^{\mu\nu} - \frac{1}{4} \gamma_{(0)}^{\mu\nu} \phi_{(1)}^2=0
\ee
we arrive at,
\ba
T^{\mu\nu} &=& 3 r^5\left(\gamma_{(3)}^{\mu\nu} - \gamma_{(0)}^{\mu\nu} \text{tr} \gamma^{(3)} -\frac{1}{3} \gamma_{(0)}^{\mu\nu}\phi_{(1)}\phi_{(2)} \right)\\
 &=& 3 r^5\left(g_{(3)}^{\mu\nu} - g_{(0)}^{\mu\nu} \text{tr} g^{(3)} - \frac{2}{3} g_{(0)}^{\mu\nu} g^{(3)}_{zz} - \frac{1}{3}g_{(0)}^{\mu\nu}\phi_{(1)}\phi_{(2)}\right)
\ea
from which we identify the expectation value of the field theory stress tensor,
\be
\left<T^{\mu\nu}\right> = \frac{1}{r^5} T^{\mu\nu} =  3 \left(g_{(3)}^{\mu\nu} - g_{(0)}^{\mu\nu} \text{tr} g^{(3)} - \frac{2}{3} g_{(0)}^{\mu\nu} g^{(3)}_{zz}-\frac{1}{3} g_{(0)}^{\mu\nu}\phi_{(1)}\phi_{(2)}\right)
\ee
Similarly we may compute the one point function for the scalar,
\be
\left<O_\phi\right> = \frac{1}{r^2}\left(\frac{1}{\sqrt{-\gamma}}\frac{\delta S_b}{\delta\phi}+\frac{2}{\sqrt{-\gamma}}\frac{\delta S_{ct}}{\delta\phi} \right) =  \phi_{(2)}
\ee
and the current,
\be
\left<J^\mu\right> = \frac{1}{r^3}\frac{1}{\sqrt{-\gamma}} \frac{\delta S}{\delta A_\mu} =  A_{(1)}^\mu.
\ee
In total the one point functions are summarised by the variation, 
\be
\delta S_{ren} = \int d^3x \sqrt{-\gamma_{(0)}}\left(\frac{1}{2}\left<T^{\mu\nu}\right>\delta \gamma^{(0)}_{\mu\nu} + \left<O_\phi\right>\delta\phi_{(1)}+ \left<J^\mu \right> \delta A^{(0)}_\mu\right).
\ee
Invariance under the Weyl transformations $\delta \gamma^{(0)}_{\mu\nu} = -2 \lambda \gamma^{(0)}_{\mu\nu}$, $\delta \phi_{(1)} = \lambda \phi_{(1)}$ gives the conformal Ward identity,
\be
\left<T^\mu_\mu\right> =  \phi_{(1)}\left<O_\phi\right>.
\ee
by inserting the expressions for the one-point functions above we verify this holds using the equations of motion.

\section{Numerical convergence\label{convergence}}
For the DeTurck quantities, $|\xi| = \sqrt{\xi^M\xi_M}$ and $|\psi|$ as defined in section \ref{gaugefix} we compute their maximum values on the grid, denoted by $|\xi |_{\text{max}}$ and $|\psi|_{\text{max}}$ respectively. In addition we wish to study how the numerically extracted free energy converges with $N$. Denoting the value of $\Delta \omega$ obtained at $N^3$ grid points as $\Delta\omega(N)$, we compute,
\be
\omega_N = \log_{10} \left|\frac{\Delta \omega(N+2^{1/3})}{\Delta \omega (N-2^{1/3})}-1\right|. \label{omegaN}
\ee

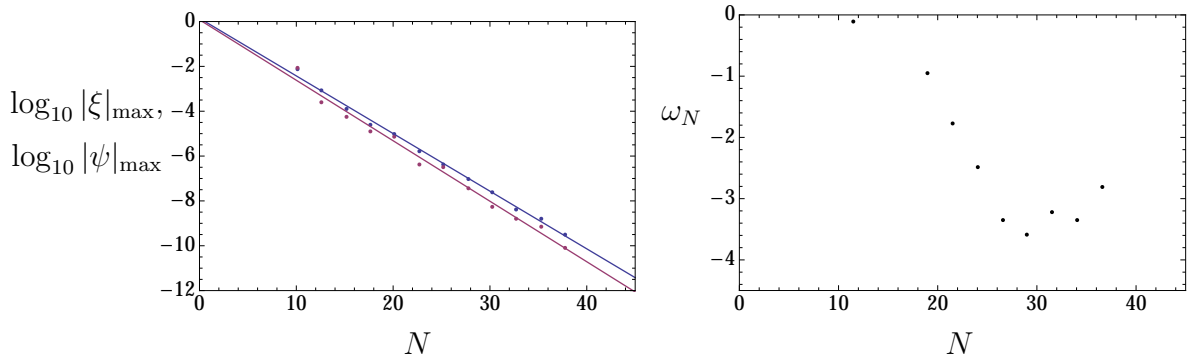
\begin{figure}[h!]
\begin{center}
\includegraphics[width=0.40\columnwidth]{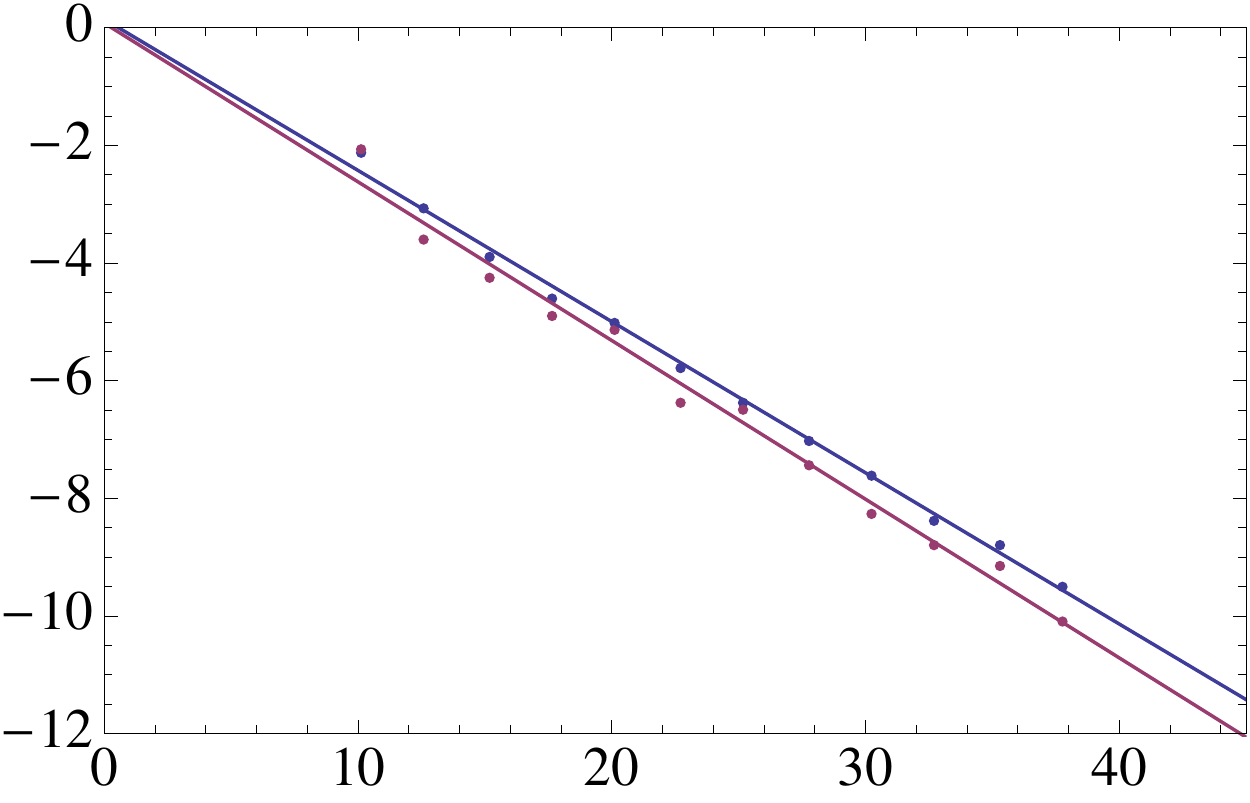}
\hspace{20pt}
\includegraphics[width=0.40\columnwidth]{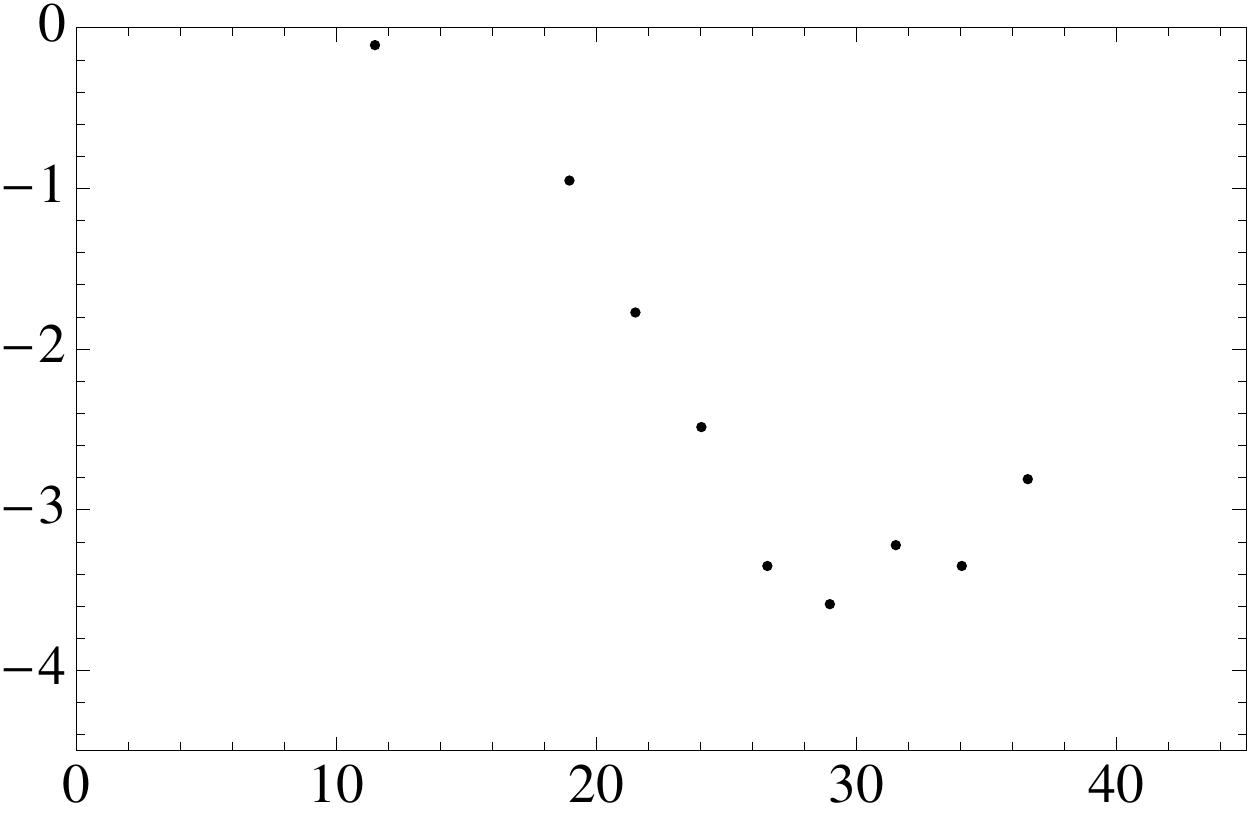}
\begin{picture}(0.1,0.1)(0,0)
\put(-420,80){\makebox(0,0){$\log_{10} |\xi |_{\text{max}},$}}
\put(-420,60){\makebox(0,0){$\log_{10} |\psi|_{\text{max}}$}}
\put(-195,76){\makebox(0,0){$\omega_N$}}
\put(-295,-10){\makebox(0,0){$N$}}
\put(-90,-10){\makebox(0,0){$N$}}
\end{picture}
\vskip 1em
\caption{Convergence with the number of grid points for the $c_1=9.9, n=36$ checkerboard at $T=0.55T^\ast$ and $k_x=k_y=k^\ast$.  The total number of grid points is $N^3$. \emph{Left panel}: Exponential convergence of gauge fixing variables towards zero. $|\xi |_{\text{max}}$ is shown in blue and $|\psi|_{\text{max}}$ in red, and are defined in the text. Also shown are the best-fit straight lines. \emph{Right panel}: The convergence of the free energy difference, $\Delta \omega$, as defined in \eqref{deltaomega} demonstrated using $\omega_N$ as defined in \eqref{omegaN}.\label{prelimconvergence}}
\end{center}
\end{figure}

In Figure \ref{prelimconvergence} we present the convergence with the number of grid points for the solution presented in section \ref{prelim}. The values of $|\xi|_{\text{max}}$ and $|\psi|_{\text{max}}$ converge towards zero exponentially. The accuracy to which we can extract $\omega_N$ (we use three numerical derivatives in this case) saturates, but at that point $\omega_N$ is very small.

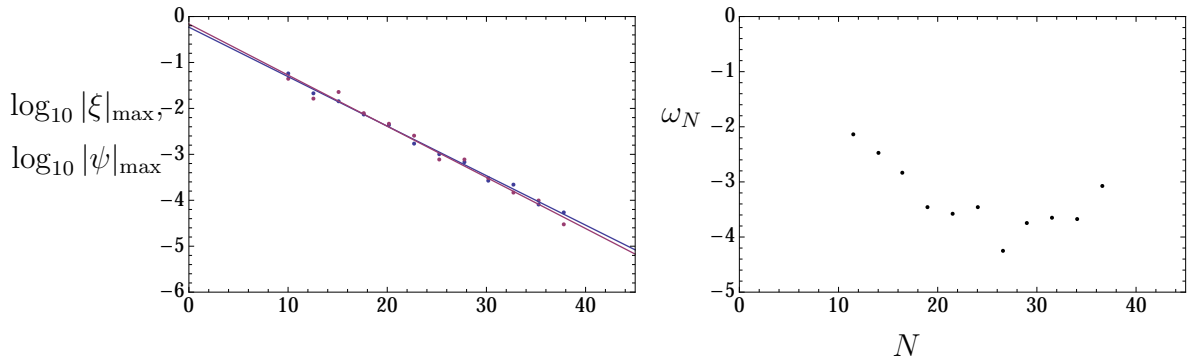
\begin{figure}[h!]
\begin{center}
\includegraphics[width=0.40\columnwidth]{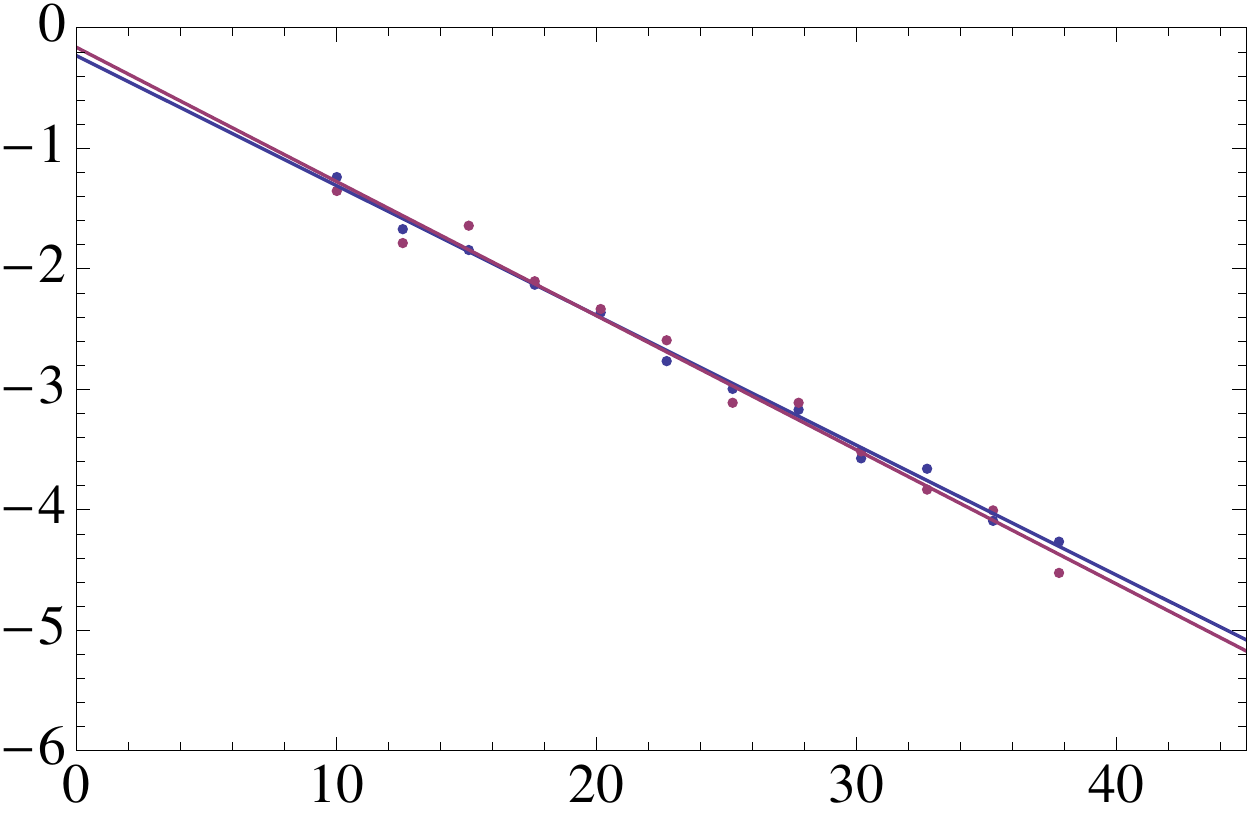}
\hspace{20pt}
\includegraphics[width=0.40\columnwidth]{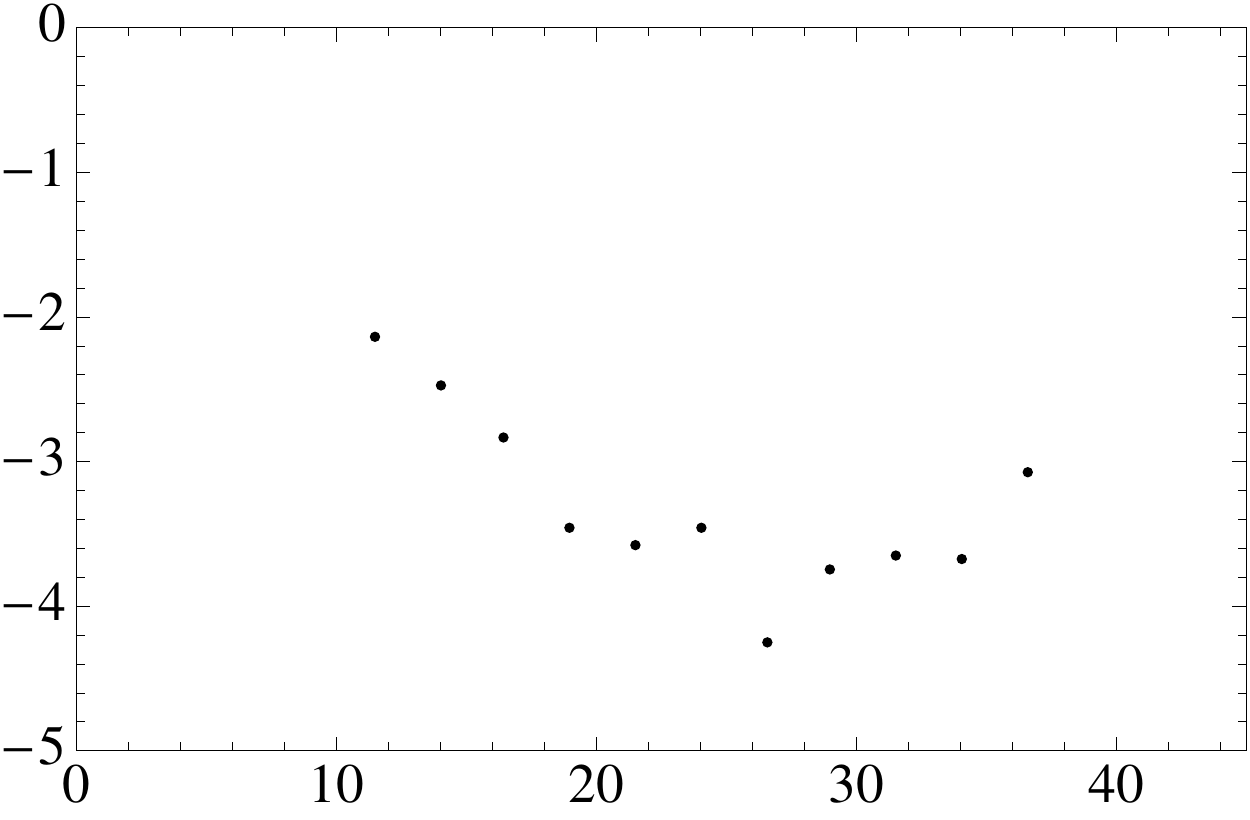}
\begin{picture}(0.1,0.1)(0,0)
\put(-420,80){\makebox(0,0){$\log_{10} |\xi |_{\text{max}},$}}
\put(-420,60){\makebox(0,0){$\log_{10} |\psi|_{\text{max}}$}}
\put(-195,76){\makebox(0,0){$\omega_N$}}
\put(-110,-10){\makebox(0,0){$N$}}
\end{picture}
\vskip 1em
\caption{Convergence with the number of grid points (for definitions see Figure \ref{prelimconvergence}) for the data of section \ref{transition} near the checkerboard-stripe first order transition, at a temperature of $T=0.04$ and momenta $k_x=k_y=0.6$. \label{convergence13}}
\end{center}
\end{figure}

In Figure \ref{convergence13} we present the convergence with the number of grid points for the data of section \ref{transition} near the checkerboard-stripe first order transition, at a temperature of $T=0.04$ and momenta $k_x=k_y=0.6$. Comparing with the other model in Figure \ref{prelimconvergence}  the values of $|\xi|_{max}$ and $|\psi|_{max}$ are not quite as small. However, they still converge towards zero exponentially fast with $N$ as required. Moreover, the free energy has converged sufficiently at the values of $N$ shown. The situation improves at higher temperatures, e.g. near the swallowtail where the values once more mirror those of Figure \ref{prelimconvergence}.

\bibliographystyle{utphys}
\bibliography{stripes}{}
\end{document}